\documentclass[%
aip,
amsmath,amssymb,
reprint,%
]{revtex4-1}

\usepackage{xcolor}
\usepackage{graphicx}
\usepackage{dcolumn}
\usepackage{bm}

\usepackage[utf8]{inputenc}
\usepackage[T1]{fontenc}
\usepackage{mathptmx}

\newcommand{\comment}[1]{}

\begin{document}
	
	\preprint{AIP/123-QED}
	
	\title[]{Controllable and Non-Dissipative Inertial Dynamics of Skyrmion in a Bosonic Platform}
	
	\author{Imam Makhfudz}
	\affiliation{IM2NP, UMR CNRS 7334, Aix-Marseille Universit\'{e}, 13013 Marseille, France}

	\date{\today}
	
	\begin{abstract}
		It has been understood in the past a decade or two that the dynamics of spin or magnetization in ultrafast regime necessarily involves inertial term that reflects the reluctance to follow abrupt or sudden change in the spin or magnetization orientation. The role of inertial spin dynamics in governing the motion of Skyrmion, a topological spin texture, is elucidated. Using nonequilibrium Green's function Keldysh formalism, an equation of motion is derived in terms of collective coordinates for a Skyrmion coupled via a ``minimal coupling'' to a bath of harmonic oscillators of frequency $\omega$, modeling an optical phonon-like bosonic bath with its nearly-flat energy spectrum, and a coupling to the phonon energy density that dominates under resonance condition at the optical phonon frequency. A deterministic and non-dissipative dynamics equation of motion is obtained with an explicit mass term for the Skyrmion emerging due to the coupling, even within rigid Skyrmion picture. This results in a cyclotronic motion of Skyrmion, with a frequency that can go ultrafast, depending on that of the oscillator. Controlling the oscillator frequency can therefore guide the Skyrmion dynamics. Our theory bridges inertial dynamics and topology in magnetism and opens a pathway to ultrafast control of topological spin textures. 
	\end{abstract}
	
	\maketitle
	

	\textit{Introduction.\textemdash}Topological spin textures such as magnetic vortices and Skyrmion offer great potential to be carrier of information for future technology \cite{FertReview}. It is therefore crucial to develop capability to control the dynamics of topological spin textures. Until recently, a Skyrmion has been modeled as massless but charged rigid object which would undertake circular motion when subject to harmonic potential and gyrotropic (Magnus) force. Numerical simulations and experimental works however reported nontrivial dynamics of magnetic bubble, a spin tecture topologically equivalent but of much larger size than Skyrmion \cite{Moutafis}, which can be described using the paradigmatic Thiele's equation \cite{ThielePRL} only if we include a mass or inertia term for the Skyrmion \cite{MakhfudzPRL}. The origin of this mass has been an object of intensive studies the past decade. This mass effect is present even within slow adiabatic dynamics and has been proposed to arise from the geometric phase effect intrinsic due to the noncollinear spin texture of the Skyrmion \cite{MakhfudzPRL}, as well as from the coupling of the spins forming the Skyrmion to other degrees of freedom constituting the environment \cite{GalitskiPRB}\cite{GaraninPRB}\cite{BKNikolicJPSJ}. The question of Skyrmion mass became contentious issue when it was later found that Skyrmion can nevertheless be massless \cite{BrinkPRB}\cite{WuTchernyshyovSciPost}. At about the same time, an interesting development occurred, where Skyrmion has been noted to display quantum properties, such as the analog of macroscopic tunneling and quantized energy level \cite{LinBulaevskiiPRB}. Skyrmion mass shows up in its quantum dynamics, demonstrated theoretically and computed using functional integral formalism \cite{PsaroudakiPRX}. The quantum behavior of Skyrmion has opened up a prospect for its application as a qubit, leading to the proposal for the so-called Skyrmion qubit \cite{SkyrmionQubitPsaroudakiPRL}.
	
	In the past fifteen years or so, it has been realized that spin or magnetization has intrinsic inertia which manifests in the ultrafast, short time scale dynamics \cite{WegrowePRB,OliveAPL,OliveJAP,BonettiNatPhys,WegroweAJP}. For example, a spin excited by femtosecond laser pulse cannot instantly follows the sudden change in the field strength of the latter, as Landau-Lifshitz equation predicts, even in the absence of damping. Rather, the spin takes some time to adjust itself to the change in the field strength, which reflects its laziness or inertia to exhibit abrupt time dependence. In terms of equation of motion, this appears as a term involving a second time derivative of the spin vector, which is relevant only on time scale shorter than the so-called inertial time. The implication of this inertial term has been demonstrated to be rich, giving rise to spin nutation\cite{WegrowePRB} that has been observed experimentally\cite{BonettiNatPhys}, inertial spin wave\cite{MakhfudzAPL}\cite{WegroweAnatomyPRB}\cite{MikhailPRB}, auto-oscillation\cite{AutoOscillationPRL} and may have practical applications in ultrafast spintronics \cite{MakhfudzPRB}. The origin of this inertial effect is however more subtle question, with proposals suggesting that it arises from intrinsic non-equilibrium thermodynamics \cite{WegrowePRB}, relativistic correction \cite{Mondal}, to coupling to electronic degree of freedom \cite{Fahnle}\cite{Fransson}. The latter mechanism has been extended recently to show that spin inertia arises from coupling of spin to a bath or environment \cite{DuinePRLinertia}. A curious question is whether spin inertia can be the origin of Skyrmion inertia, when thousands of spins are excited out of ferromagnetic ground state to form a Skyrmion. 
	
	In this Letter, we ask the question of, and provide the answer for it, how spin inertia would affect the dynamics of a topological spin texture, Skyrmion more specifically, when the Skyrmion is assumed to be coupled to a bosonic platform made of harmonic oscillators. The platform induces inertia for each spin and since the Skyrmion is made up of many spins, the platform eventually induces and controls an inertial dynamics of the Skyrmion. In fact, the harmonic oscillators drive the Skyrmion dynamics into ultrafast regime.  This system is physically relevant; it applies to the Skyrmion in magnets with significant magnetoelastic coupling, where the spins are coupled to the lattice or elastic degree of freedom.     
	
	\textit{Model of Skyrmion Coupled to Bosonic Environment.\textemdash}In many magnets, the magnetization is produced by the spins localized at the lattice sites. A two-dimensional lattice of spins hosting a single Skyrmion, which then couples to the vibration of the lattice quantized in terms of phonons, is illustrated in Fig. 1.  
	\begin{figure}
		\includegraphics[angle=0,origin=c, scale=0.30]{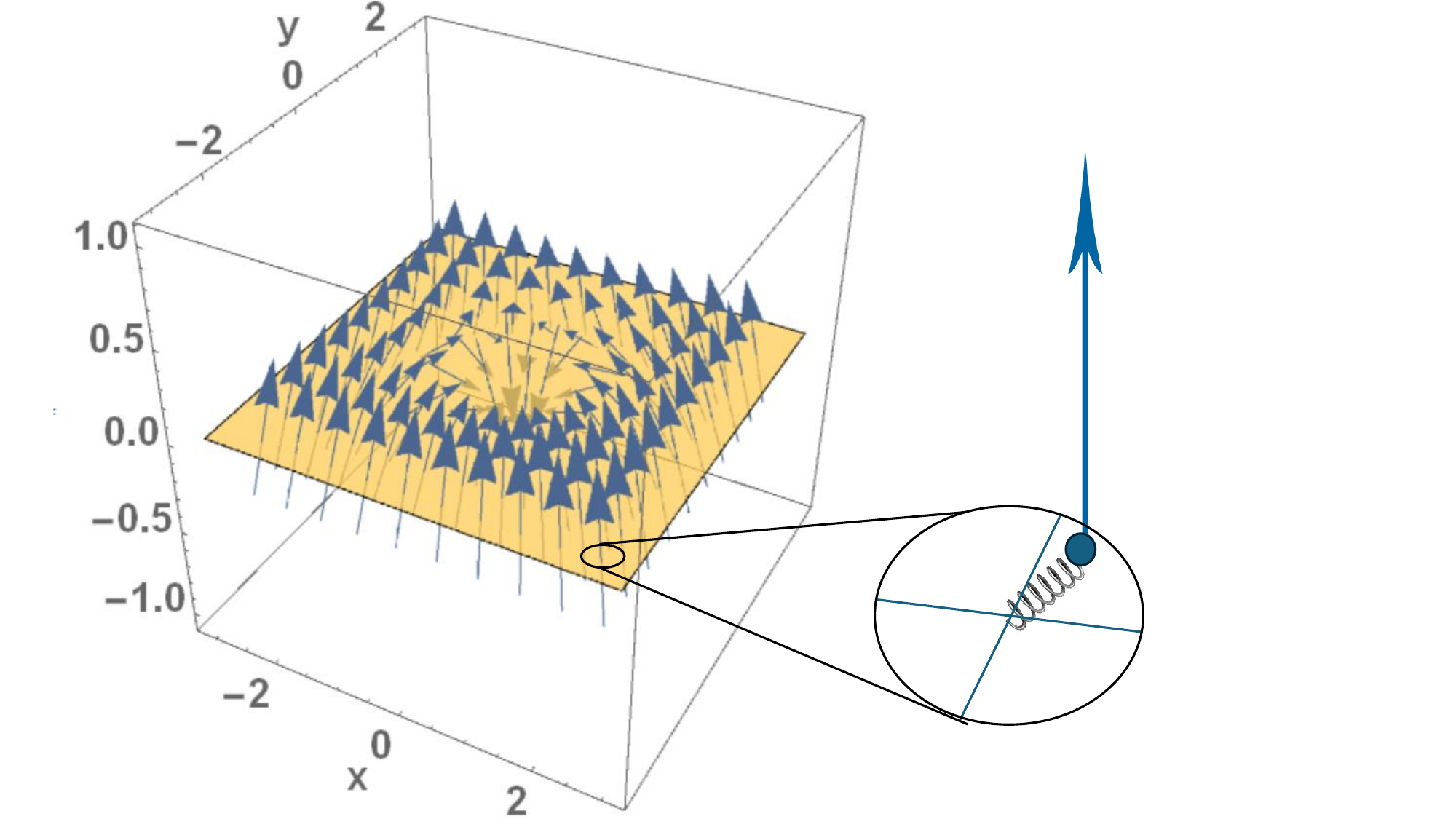}
		\caption{
			Illustrative profile of a Skyrmion on a two-dimensional (2D) lattice (arylide yellow plane). The spins sit on a lattice whose vibration is modeled by harmonic oscillators whose displacement field is coupled to the Skyrmion. The axes are labeled in dimensionless coordinates; $x=\tilde{x}/(2a),y=\tilde{y}/(2a)$ where $\tilde{x},\tilde{y}$ are dimensioned coordinates while $a$ is the lattice spacing. The inset schematically illustrates the coupling between the spin and the lattice harmonic oscillator, modeled as a bob carrying spin (arrow) connected to the spring attached to the lattice site.}\label{fig:Fig1}
	\end{figure}
	The system is described the following Hamiltonian,
	\begin{equation}\label{ModelHamiltonian}
		H=H_{Sk}+H_B+H_{B-Sk}
	\end{equation}
	where $H_{Sk}$ describes the energy of the magnetic degree of freedom, which hosts Skyrmion solution. This magnetic energy Hamiltonian can be taken to be any Hamiltonian that admits a single Skyrmion solution, characterized by Skyrmion number $N_{Sk}=(4\pi)^{-1}\int d^2\mathbf{r}\mathbf{S}\cdot\left(\partial_x\mathbf{S}\times\partial_y\mathbf{S}\right)=-n$, where $n$ is the vorticity \cite{LinHayamiPRB} in the parameterization of the spin $\mathbf{S}=(\sin\Theta\cos\Phi,\sin\Theta\sin\Phi,\cos\Theta)$, where $\Theta=\Theta(r),\Phi=n\phi+\varphi$ with $\varphi$ is the helicity. For concreteness, we take the magnetic Hamiltonian to be that of non-chiral magnets\cite{LinHayamiPRB} 
	\begin{equation}\label{HSk}
		H_{Sk}=-\frac{I_1}{2}\left(\nabla \mathbf{S}\right)^2+\frac{I_2}{2}\left(\nabla^2 \mathbf{S}\right)^2-\mathbf{H}_a\cdot\mathbf{S}
	\end{equation}    
	with constraint $|\mathbf{S}|=1$, $I_1$ and $I_2$ are the exchange energies that set the length scale for the system, while $\mathbf{H}_a$ is a static magnetic field \cite{LinHayamiPRB}. The first, second, and third terms respectively represent the quadratic exchange, quartic exchange, and Zeeman energies. \textcolor{black}{With $I_1,I_2>0$, the quadratic term in $\nabla \mathbf{S}$ prefers noncollinear order while the quartic term helps stabilize the system. This is an effective Hamiltonian describing spin interactions in inversion-symmetric magnets, where Dzyaloshinskii-Moriya interaction, common in chiral magnets where inversion symmetry is broken, is now forbidden by symmetry. Different from chiral magnets, inversion symmetric magnets therefore rely on the interplay between competing spin exchange interactions on frustrated geometry such as triangular lattice and Zeeman field to stabilize the Skyrmion. For the particular class of candidate materials considered here, the frustrated competing exchange interactions on triangular lattice with effective continuum Hamiltonian described by Eq.(\ref{HSk}), lead at zero temperature to noncollinear spin configuration, a conical spiral structure more precisely, at low Zeeman field $H_a\leq H_s=1/4$ where $H_s$ is the saturation field in dimensionless unit. Increasing the Zeeman field past $H_s$ drives a continuous transition to uniform ferromagnetically polarized state. In this ferromagnetic background, a single Skyrmion solution can exist as a metastable state which, different from that in chiral magnets, has diverging size close to the critical field $H_s$\cite{LinHayamiPRB}.}
	
	Since a Skyrmion typically contains thousands of spins, we will adopt collective coordinate approach \cite{PRD1}\cite{PRD2}, where the Skyrmion is assumed to be a rigid object of fixed shape while its center of mass gets displaced, allowing us to parameterize its motion simply in terms of the position of the center of mass $\mathbf{R}(t)$; the angles will be written as $\Phi(\mathbf{r})=\Phi_0(\mathbf{r}-\mathbf{R}(t)),\Theta(\mathbf{r})=\Theta_0(\mathbf{r}-\mathbf{R}(t))$ describing the profile of rigid Skyrmion with its center of mass position vector $\mathbf{R}(t)$.   
	
	The bosonic bath Hamiltonian is that of a continuous distribution of isotropic harmonic oscillators in two dimensions
	\[
	H_B=\frac{1}{2}\rho\dot{\mathbf{u}}^2+\frac{1}{2}\rho\omega^2\mathbf{u}^2=\frac{\mathbf{p}^2}{2\rho}+\frac{1}{2}\rho\omega^2\mathbf{u}^2
	\]
	\begin{equation}\label{BathHamiltonian}
		=\hbar\omega\sum_{\alpha=x,y} \left(b^{\dag}_{\alpha}b_{\alpha}+\frac{1}{2}\right)
	\end{equation}
	where $\rho$ is the mass area density of the oscillator, $\mathbf{u}$ is the oscillator displacement vector field $\mathbf{u}=\mathbf{u}(\mathbf{r},t)=(u_x(\mathbf{r},t),u_y(\mathbf{r},t))$, and $\omega$ is the oscillator frequency, and $\mathbf{p}=\rho\dot{\mathbf{u}}$ is the momentum conjugate to $\mathbf{u}$. Standard quantization of the harmonic oscillator ($u_{\alpha}(\mathbf{r})=\sqrt{\hbar/(2\rho\omega)}(b^{\dag}_{\alpha}(\mathbf{r})+b_{\alpha}(\mathbf{r})),p_{\alpha}(\mathbf{r})=i\sqrt{\hbar\rho\omega/2}(b^{\dag}_{\alpha}(\mathbf{r})-b_{\alpha}(\mathbf{r}))$) gives the last line of Eq.(\ref{BathHamiltonian}), with $b^{\dag}_{\alpha}(\mathbf{r}),b_{\alpha}(\mathbf{r})$ respectively the boson creation and annihilation operators. Instead of direct linear coupling between the spin (or magnetization density $\mathbf{n}$) and the bath degrees of freedom $\mathbf{p},\mathbf{u}$ which is not evident since the Skyrmion contains thousands of spins that are non-collinear, we consider a ``minimal but natural coupling'' directly in terms of collective coordinates, by setting $\mathbf{p}(\mathbf{r})\rightarrow\mathbf{p}(\mathbf{r}-\mathbf{R}(t)),\mathbf{u}(\mathbf{r})\rightarrow\mathbf{u}(\mathbf{r}-\mathbf{R}(t))$. Upon quantization, we will have $b^{\dag}_{\alpha}(\mathbf{r})\rightarrow b^{\dag}_{\alpha}(\mathbf{r}-\mathbf{R}(t)), b_{\alpha}\rightarrow b_{\alpha}(\mathbf{r}-\mathbf{R}(t))$. The resulting Hamiltonian then encodes the bath and its coupling to the Skyrmion. Expanding this Hamiltonian to linear order in $\mathbf{R}$, we obtain  
	\begin{equation}\label{Bath+BathSkyrmionHamiltonian}
		H_{B+B-Sk}=\hbar\omega\sum_{\alpha=x,y} \left(b^{\dag}_{\alpha}b_{\alpha}+\frac{1}{2}\right)
		-\hbar\omega\sum_{\alpha=x,y}R_i\frac{\partial }{\partial r_i}\left(b^{\dag}_{\alpha}b_{\alpha}\right)
	\end{equation}
	which will be used as the starting point of the derivation of the corresponding action for the bath and its coupling to the Skyrmion. One can view this model as involving a 
	nonlinear coupling between Skyrmion and the harmonic oscillators\cite{Note}. This phenomenological coupling is best viewed as a coupling of the (rigid) Skyrmion to the energy density of the harmonic oscillator lattice. In this case, the harmonic oscillators are independent from each other (i.e. decoupled), which is simpler than phonon Hamiltonian in periodic solids, where the harmonic oscillators are coupled to each other. As established in the literature, in magnetic (periodic) solids, the lattice is normally coupled to the magnetic degree of freedom via magnetoelastic coupling\cite{MEc1}\cite{MEc2}\cite{MEc3}. The principal feature of our minimal coupling is that it is quadratic in the oscillator displacement field $\mathbf{u}$ while leading order term in the familiar magnetoelastic coupling is linear in $\mathbf{u}$, both of which however are allowed by symmetry \cite{KaganovTsukernik}. The main results of this work only rely on this principal feature, with details given in the last section of \cite{SupplementaryMaterials}. 
	
	Since the Hamiltonian Eq.(\ref{BathHamiltonian}) corresponds to independent harmonic oscillators, in the Fourier space, it can be written as $H_B= \sum_{\mathbf{k}\alpha}\hbar\omega_{\mathbf{k}} \left(b^{\dagger}_{\mathbf{k}\alpha}b_{\mathbf{k}\alpha}+1/2\right)$
	where $\omega_{\mathbf{k}}=\omega$. So, a system of independent (uncoupled) harmonic oscillators has a flat energy dispersion, i.e. dispersionless spectrum. One can therefore view our bosonic system as the approximate model for optical phonon mode in periodic solids because its spectrum is nearly flat \cite{BornHuangBook}. Detailed derivation indicates that our minimal coupling leads to a dynamical term (cf. the very last term in Eq.(\ref{SphiR}), the form of which is universal) that always accompanies the the standard magnetoelastic coupling \cite{RuckriegelPRB} within the Lagrangian or action formulation. The more accurate phonon Hamiltonian and standard magnetoelastic coupling lead to an effective action that reduce to that of our simple model, in the limit of large distance and or large Skyrmion size, justifying our model\cite{SupplementaryMaterials}. While the standard magnetoelastic coupling acts on both the acoustic and optical phonons, under resonance condition; i.e. when phonons are excited only at or near the optical phonon frequency, the dynamical coupling to the optical phonon prevails, consistent with the general understanding that spin-lattice coupling is maximized at resonance\cite{MEc3}.
	
	Following the standard form of the action for a spin \cite{AuerbachBook}, the action for the Skyrmion can be written as\cite{NoteNotation}
	\begin{equation}\label{SkyrmionAction}
		S_{Sk}=\int dt \int d^2\mathbf{r} \left(\mathbf{A}\cdot \dot{\mathbf{S}}-\mathcal{H}\left[\mathbf{S}\right]\right)
	\end{equation}
	where $\mathbf{A}$ is the monopole vector potential, a gauge-dependent quantity, while $\mathcal{H}[\mathbf{S}]$ is the magnetic energy density given based on Eq.(\ref{HSk}) where the vector $\mathbf{S}$ now takes a Skyrmion configuration as illustrated by Fig.~\ref{fig:Fig1}. Rewriting the monopole part $S^{\mathbf{A}}_{Sk}=\int dt\int d^2\mathbf{r}\mathbf{A}\cdot\dot{\mathbf{S}}$ explicitly in terms of the angles spin parameterization, it becomes
	\begin{equation}\label{SkyrmionActionA}
		S^{\mathbf{A}}_{Sk}=\rho_s\int dt \int d^2\mathbf{r}\dot{\Phi}_0\left(\mathbf{r}-\mathbf{R}(t)\right)\left(1-\cos\Theta_0\left(\mathbf{r}-\mathbf{R}(t)\right)\right)
	\end{equation}
	where $\rho_s$ is the density of spin angular momentum (per unit area). We have chosen a specific gauge $\mathbf{A}=\left(1-\cos\Theta\right)/\sin\Theta\hat{\phi}$, but the resulting classical equation of motion to be derived later is independent of the gauge choice. Expanding the angle variables, $\Theta_0(\mathbf{r}-\mathbf{R}(t))=\Theta_0(\mathbf{r})-\mathbf{R}\cdot\nabla\Theta(\mathbf{r}),\Phi_0(\mathbf{r}-\mathbf{R}(t))=\Phi_0(\mathbf{r})-\mathbf{R}\cdot\nabla\Phi_0(\mathbf{r})$ we find to lowest order terms,
	\begin{equation}\label{MonopoleActionCC}
		S^{\mathbf{A}}_{Sk}=S_{\mathrm{topol}}+S_{\mathrm{WZW}}
	\end{equation}
	where
	\begin{equation}
		S_{\mathrm{topol}} = 4\pi N_{Sk}\rho_s \int \left(R_x\dot{R}_y-R_y\dot{R}_x\right)dt 
	\end{equation}
	and
	\begin{equation}
		S_{\mathrm{WZW}}=-\rho_s\int dt\int d^2\mathbf{r}\dot{\mathbf{R}}\cdot\left(1-\cos\Theta_0\right)\nabla\Phi_0
	\end{equation}
	with $N_{Sk}=(4\pi)^{-1}\int d^2\mathbf{r}\sin\Theta_0(\partial_x\Theta_0\partial_y\Phi_0-\partial_y\Theta_0\partial_x\Phi_0)$ is the Skyrmion number in terms of the angular fields and we have implied $\mathbf{R}=(R_x, R_y)\equiv (X,Y)$ and $S=1$. The topological action $S_{\mathrm{topol}}$ was noted in early work \cite{MStonePRB} while the Wess-Zumino-Witten (WZW) action may give rise to quantum phenomena $S_{\mathrm{WZW}}$ \cite{BraunLoss}\cite{LossDiViGrinPRL}. This WZW term vanishes for most Skyrmion solutions of interest, including the non-chiral Skyrmion. 
	
	The magnetic energy part will be given by that for a single Skyrmion solution \cite{LinHayamiPRB},
	\begin{equation}
		\mathcal{H}\left[\mathbf{S}\right]=E_T=E_2+E_4-H_a\cos\Theta
	\end{equation}
	where each of the terms in the energy density $E_T$ is function of $\Theta(r)$, which only depends on distance $r$ from the center of Skyrmion (taken to be at the reference origin of coordinate for static Skyrmion in equilibrium). The profile of the Skyrmion can be specified in terms of $\theta(r)$ obtained by minimizing $\mathcal{E}_T$ with respect to $\theta$ \cite{LinHayamiPRB}, satisfying $\partial E_T/\partial \Theta=0$. The magnetic energy density part in the action written in terms of collective coordinates of the Skyrmion becomes
	\begin{equation}\label{SkyrmionActionMagnetic}
		S^M_{Sk}=\int dt \int d^2\mathbf{r} \left(-\mathcal{H}\left[\mathbf{S}\right]\right)=S_0+\int dt \left(\mathbf{F}_2+\mathbf{F}_4+\mathbf{F}_H\right)\cdot \mathbf{R}
	\end{equation}
	where $S_0=-\int dt\int d^2\mathbf{r}E_T$ a constant action and the force $\mathbf{F}_2=\int d^2\mathbf{r}(\partial E_2/\partial\mathbf{r})$ and similarly for the other two forces, which are independent of $\mathbf{R}$. In equilibrium, the net force of the rigid Skyrmion is zero.

	\textit{Effective Action of Skyrmion.\textemdash}We write the full action of the system of Skyrmion and harmonic oscillators and integrate out the latter to get an effective action for the Skyrmion\cite{SupplementaryMaterials}. Following the procedure to derive the action for bosons through evaluation of the partition function in terms of the time evolution operator along closed time contour of Keldysh \cite{KamenevBook}, the action of the boson of the bath coupled to the Skyrmion is found to be
	\[
	S[\overline{\phi},\phi,\mathbf{R},\dot{\mathbf{R}}]=\int dt \int d^2\mathbf{r}\sum_{\alpha}\overline{\phi}_{\alpha}(\mathbf{r},t)\hbar\left(i\partial_t-\omega\right)\phi_{\alpha}(\mathbf{r},t)
	\]
	\begin{equation}\label{SphiR}
		+\int dt \int d^2\mathbf{r}\hbar\omega\sum_{\alpha}\left(\mathbf{R}\cdot\nabla|\phi_{\alpha}(\mathbf{r},t)|^2-\overline{\phi}_{\alpha}(\mathbf{r},t)\left(\tau\dot{\mathbf{R}}\cdot i\nabla\right)\phi_{\alpha}(\mathbf{r},t)\right)
	\end{equation}   
	where $\phi_{\alpha}$ is the bosonic coherent state for species $\alpha=x,y$ and $\overline{\phi}_{\alpha}$ is its conjugate, obeying $b_{\alpha}|\phi_{\alpha}\rangle=\phi_{\alpha}|\phi_{\alpha}\rangle,\langle \overline{\phi}_{\alpha}|b^{\dag}_{\alpha}=\langle \overline{\phi}_{\alpha}|\overline{\phi}_{\alpha}$. The first term in Eq.(\ref{SphiR}) is standard for the action of bosons with Hamiltonian of the form $b^{\dag}b$ while the remaining two last terms represent the effect of coupling to Skyrmion. The third term involves a time scale $\tau$ that is characteristic time scale of the Skyrmion dynamics, that is distinct from that of the bath harmonic oscillator dynamics. In fact\cite{SupplementaryMaterials}, the $\tau$ can be associated with the time scale that separates the inertial regime ($t<\tau$) from the adiabatic regime ($t>\tau$) in the spin dynamics described by inertial Landau-Lifshitz-Gilbert equation \cite{WegrowePRB}\cite{OliveAPL}\cite{OliveJAP}. The contribution of the second term eventually vanishes because of the total gradient. The coupling constant in the last term is proportional to the dimensionless parameter $\omega\tau$, permitting a perturbation theory in appropriate regimes of parameter space when this parameter is smaller than unity. The overall dynamical Skyrmion-lattice coupling constant in our theory is thus proportional to the product of the optical phonon frequency $\omega$ and the inertial time $\tau$, the latter of which depends on the strength of the magnetocrystalline anisotropy \cite{BonettiPRL}.
	
	We then work in Fourier space for the spatial part,
	\begin{equation}
		\phi_{0\alpha}(\mathbf{r},t)=\frac{1}{\sqrt{A}}\sum_{\mathbf{k}}\phi_{0\mathbf{k},\alpha}(t)e^{i\mathbf{k}\cdot\mathbf{r}}\equiv\sqrt{A}\int \frac{d^2\mathbf{k}}{(2\pi)^2}\phi_{0\mathbf{k},\alpha}(t)e^{i\mathbf{k}\cdot\mathbf{r}}
	\end{equation}
	where $A=L_xL_y$ is the area of the 2-dimensional lattice and $\phi_{0\mathbf{k},\alpha}(t)$ is the corresponding $\mathbf{k}$ space coherent state field. Redefining the coherent state fields and Skyrmion collective coordinates along the forward ($+$) and backward ($-$) parts of Keldysh closed-time contour and performing rotation to the classical ($cl$) and quantum ($q$) versions $\phi^{\pm}_{0\alpha}=\left(\phi^{cl}_{0\alpha}\pm\phi^q_{0\alpha}\right)/\sqrt{2}$, $\mathbf{R}^{\pm}=\mathbf{R}_c\pm\mathbf{R}_q/2$ applied to Eq.(\ref{BathHamiltonian}) gives a bath+bath-Skyrmion Keldysh action\cite{SupplementaryMaterials}
	\[
	S_{B+B-Sk}=\int^{+\infty}_{-\infty}dt \int^{+\infty}_{-\infty}dt'A\int \frac{d^2\mathbf{k}}{(2\pi)^2}\overline{\mathbf{V}}^T(\mathbf{k},t)
	\]
	\begin{equation}\label{bathSkyrmionaction}
		\times
		\begin{pmatrix}
			[\tilde{G}^{-1}]^{cl,cl} & [G^{-1}]^A + [\tilde{G}^{-1}]^{cl,q}\\
			[G^{-1}]^R + [\tilde{G}^{-1}]^{q,cl}& [G^{-1}]^K + [\tilde{G}^{-1}]^{q,q}
		\end{pmatrix}_{t,t'}	
		\mathbf{V}(\mathbf{k},t')
	\end{equation}
	where we have defined a 2-element column vector of bosonic fields $\overline{\mathbf{V}}^T(\mathbf{k},t)=(\overline{\phi}^{cl}_0 (\mathbf{k},t), \overline{\phi}^{q}_0(\mathbf{k},t))$ and $\tilde{\mathbf{V}}(\mathbf{k},t)=(\phi^{cl}_0 (\mathbf{k},t), \phi^{q}_0(\mathbf{k},t))$ with $T$ indicates transpose. The retarded ($R$), advanced ($A$), and Keldysh ($K$) Green functions corresponding to the pure bath action are given by
	\begin{equation}
		[G^{-1}]^{A(R)}(t,t') = \hbar\left(i\partial_t-\omega\mp i0\right)\delta(t-t')
	\end{equation}
	\begin{equation}\label{KeldyshGF}
		[G^{-1}]^K (t,t')=2i0 F(t,t')
	\end{equation}
	where $F(t,t')$ is a Hermitian matrix. On the other hand, the Green functions corresponding to bath-Skyrmion coupling action are
		\begin{equation}\label{pureclassicquantumparts}
			[\tilde{G}^{-1}]^{cl,cl}(t,t') = 0,  [\tilde{G}^{-1}]^{q,q}(t,t')= 2i 0 F(t,t') 
		\end{equation}
		\begin{equation}
			[\tilde{G}^{-1}]^{cl,q(q,cl)}(t,t') = \hbar\omega\tau\dot{\mathbf{R}}_c\cdot\mathbf{k}\delta(t-t')
		\end{equation}
	where we have retained only the zeroth order terms in $\dot{\mathbf{R}}_q$, which is indeed a small (quantum fluctuation) quantity $|\dot{\mathbf{R}}_q|\ll |\dot{\mathbf{R}}_c|$. This simplification neglects stochasticity of harmonic oscillator reservoir (coupled to $\dot{\mathbf{R}}_q$), but justified in the classical limit of Skyrmion motion and eventually opens a road to controllable dynamics of the latter, where the Skyrmion is ``driven'' by the oscillators. The $\overline{\phi}^q_0(\mathbf{k},t)(\cdots)\phi^q_0(\mathbf{k},t)$ term in Eq.(\ref{bathSkyrmionaction}) will eventually drop out too because the corresponding Green's function matrix element $[G^{-1}]^K + [\tilde{G}^{-1}]^{q,q}$ vanishes, as implied by Eqs.(\ref{KeldyshGF}-\ref{pureclassicquantumparts}). The Green functions explicitly contains first time derivative of the Skyrmion collective coordinates $\mathbf{R}$ in their classical version. It is to be noted that the effective action of the bath and its coupling to Skyrmion Eq.(\ref{bathSkyrmionaction}) is fully quadratic in the bosonic fields; there is no linear term in the bosonic fields $(\overline{\phi}^{cl}_0 (\mathbf{k},t), \overline{\phi}^{q}_0(\mathbf{k},t))$, which is normally present in simple model of particle linearly coupled to harmonic oscillator \cite{KamenevBook}. 
	
	Defining the partition function in terms of functional integral, 
	\begin{equation}
		Z=\int \mathcal{D}\mathbf{R}e^{iS[\mathbf{R},\dot{\mathbf{R}}]}\int\mathcal{D}[\phi]e^{iS_{B+B-Sk}[\phi,\mathbf{R},\dot{\mathbf{R}}]}
	\end{equation}
	we integrate out the bath harmonic oscillator displacement field $\mathbf{u}$ and arrive at
	\begin{equation}
		Z=\int \mathcal{D}\mathbf{R}e^{i(S[\mathbf{R},\dot{\mathbf{R}}]+\delta S[\mathbf{R},\dot{\mathbf{R}}])}
	\end{equation}
	where
	\begin{equation}
		\delta S[\mathbf{R},\dot{\mathbf{R}}]=\hbar\omega\int dtA\int\frac{d^2\mathbf{k}}{\left(2\pi\right)^2}\mathrm{ln} [\mathrm{det}\left(\left(\hbar\omega\right)^{-1} G^{-1}\right)]
	\end{equation}
	where $G^{-1}$ is the matrix appearing in the action Eq.(\ref{bathSkyrmionaction}), which should also be interpreted as matrix in $t,t'$ space. It is to be noted that there is no non-local in time kernel that normally appears in model with linear coupling \cite{KamenevBook}.  This means the resulting dynamics will be non-dissipative. \textcolor{black}{This non-dissipative property is a unique feature of the dynamics of the Skyrmion coupled quadratically to extrinsic bosonic degree of freedom as provided by the harmonic oscillator or phonon. The dynamics of Skyrmion coupled to magnon for example, is inherently dissipative \cite{PsaroudakiPRX}\cite{SkyrmionQubitPsaroudakiPRL}, because magnon corresponds to intrinsic degree of freedom (small amplitude excitation on top of rigid profile) coupled linearly to the Skyrmion collective coordinate. Another mechanism which also delivers a non-dissipative dynamics relies on gyrotropic coupling, which originates from geometric Berry phase effect associated with the Skyrmion domain wall acting as topological defect, the magnitude of which is however fixed by material's magnetic properties and is not practically controllable \cite{MakhfudzPRL}. Furthermore, this dissipationless dynamics is shown to prevail even for the more familiar type of magnetoelastic coupling, in the limit of large Skyrmion size\cite{SupplementaryMaterials}, which is realized by the Skyrmion in inversion-symmetric magnets near critical field\cite{LinHayamiPRB}.}    
	
	Analogous to the result of second-order perturbation theory in the problem of spin coupled to conduction electrons \cite{NunezDuinePRB},
	the non-vanishing leading order term is the quadratic order term in the expansion of the logarithm, giving
	\begin{equation}
		\delta S[\mathbf{R},\dot{\mathbf{R}}]=\hbar\omega \int dt A\int\frac{d^2\mathbf{k}}{(2\pi)^2}O(\mathbf{k},t;\mathbf{k},t)
	\end{equation}
	where 
	\begin{equation}\label{PerturbativeActionCorrection}   
		O(\mathbf{k},t;\mathbf{k},t)=  -\frac{1}{2}\left(\left(G^R[\tilde{G}^{-1}]^{q,cl}\right)^2+\left(G^A[\tilde{G}^{-1}]^{cl,q}\right)^2\right)
	\end{equation}
	since the linear terms in $[\tilde{G}^{-1}]^{cl,q(q,cl)}$ vanish upon trace because of periodic boundary condition in time. The first term on the right hand side of the Eq.(\ref{PerturbativeActionCorrection}) (nonlocal in time) eventually vanishes, following Eq.(\ref{pureclassicquantumparts}) when the limit to zero is taken. The final result for the correction to the Skyrmion action due to bosonic bath is 
	\begin{equation}\label{SkyrmionActionCorrection}
		\delta S[\mathbf{R},\dot{\mathbf{R}}]=A\frac{\hbar\omega\tau^2}{16\pi}\Lambda^4 \int dt \left(\dot{R}^2_x+\dot{R}^2_y\right)
	\end{equation}
	where $\Lambda=1/a$ is the ultraviolet cutoff in $\mathbf{k}$ space, where $a$ is the microscopic lattice spacing. 
	
	\textit{Deterministic Inertial Skyrmion Dynamics.\textemdash}Having obtained the effective action for the Skyrmion, the corresponding equations (EOMs) of motion for $R_x,R_y$ derived by taking the change of the total Skyrmion action $S[\mathbf{R}]+\delta S[\mathbf{R},\dot{\mathbf{R}}]$ from Eqs.(\ref{MonopoleActionCC},\ref{SkyrmionActionMagnetic},\ref{SkyrmionActionCorrection}) to be zero (the least action principle) or, equivalently, the Euler-Lagrange equation. The EOMs are found to be
	\begin{equation}\label{EOMforR}
		A\frac{\hbar\omega\tau^2}{8\pi}\Lambda^4\ddot{R}_i+8\eta_i\pi N_{Sk}\rho_s\dot{R}_{\tilde{i}}=F_{2i}+F_{4i}+F_{Hi}
	\end{equation}
	where $i=x,y$, when $\tilde{i}=y,x$ and $\eta_i=-1(+1)$ for $i=x(y)$. The EOMs are local in time and the dynamics is therefore non-dissipative. Remarkably, the Skyrmion-harmonic oscillator coupling induces a mass for the Skyrmion; i.e. the first term on the left hand side, with magnitude 
	\begin{equation}
		m_{Sk}=    A\frac{\hbar\omega\tau^2}{8\pi}\Lambda^4
	\end{equation}
	proportional to the harmonic oscillator frequency and square of characteristic time scale $\tau$ of the spin dynamics. The second term on the left hand side of Eq.(\ref{EOMforR}) represents the gyrotropic force, proportional to Skyrmion number, which is the analog of Lorentz force for charged particle; the charge being the topological charge $N_{Sk}$ in this case. The forces on the right hand side are due to magnetic energy. The proportional dependence of the Skyrmion mass on the harmonic oscillator frequency is quite intuitive and transparent; the higher the frequency by which the harmonic oscillators ``shake'' the Skyrmion, the lazier the latter follows the shake up, thus the larger its inertial mass is. 
	
	The general solution of the above equation of motion is,
	\begin{equation}\label{SolutionRx}
		R_x(t)=\frac{1}{8\pi N_{Sk}\rho_s}\sum F_y t+ c_3 -\frac{c_2\cos\Omega t}{\Omega}+\frac{c_1}{\Omega}\sin\Omega t
	\end{equation}
	\begin{equation}\label{SolutionRy}
		R_y(t)=-\frac{1}{8\pi N_{Sk}\rho_s}\sum F_x t+\frac{1}{\Omega}\left(c_1\cos\Omega t+c_2\sin\Omega t\right)
	\end{equation}
	where $\sum F_i=F_{2i}+F_{4i}+F_{Hi}$ and $c_1,c_2,c_3$ are constants to be fixed by initial condition,
	\begin{equation}\label{SkyrmionFrequency}
		\Omega = \frac{64\pi^2N_{Sk}\rho_s}{A\Lambda^4\hbar\omega\tau^2}=\frac{8\pi N_{Sk}\rho_s}{m_{Sk}}\sim \left(N_{Sk}\frac{a^4}{L^4}\frac{\rho_sA}{\hbar}\frac{1}{\xi^2}\right)\omega
	\end{equation}
	the cyclotronic frequency of Skyrmion motion driven by the harmonic oscillator bath, where $\xi=\omega\tau$ is a dimensionless parameter reflecting the ratio of the harmonic oscillator frequency to spin dynamics characteristic frequency.
	
	The above solution indicates that the rigid Skyrmion exhibits driven oscillatory motion around its static reference point with a frequency $\Omega$ whose magnitude relative to the oscillator frequency $\omega$; i.e. $\Omega/\omega$ goes inversely proportional to the square of the dimensionless perturbation parameter $\xi$. Since the latter is supposed to be small for the perturbation result Eq.(\ref{PerturbativeActionCorrection}) to be justified, the resulting Skyrmion oscillation frequency can actually be very large. This result implies that the coupling to harmonic oscillators of the bath may induce ultrafast dynamics for the Skyrmion, thus entering the latter into inertial regime. The harmonic oscillators effectively drive the bounded cyclotron  motion of the Skyrmion, with both frequency and amplitude controlled by the frequency of the driver.  It is to be noted that in the $\omega=0$ limit, the harmonic oscillators are decoupled from the Skyrmion, the mass term in Eq.(\ref{EOMforR}) perishes, and the cyclotronic motion disappears.
	
	The whole term in parentheses in Eq.(\ref{SkyrmionFrequency}), obtained by taking $L_x=L_y=L$, is a dimensionless factor that can be made large for ferromagnets with large spin density and small $\xi$. Making a rough estimate using realistic parameters with $N_{Sk}=1,\rho_s=\hbar/a^2,L=1000a, a=0.55$nm, corresponding to a sample size of 0.55$\mu m \times$ 0.55$\mu m$, we obtain the dependence of the Skyrmion frequency $\Omega$ on the oscillator frequency $\omega$ and characteristic spin dynamics time scale $\tau$ as shown in Fig. 2. Low-frequency harmonic oscillators induce high frequency Skyrmion dynamics, that can go beyond GHz regime. If we take typical acoustic phonon frequency as harmonic oscillator frequency, $\omega=10$GHz and $\tau=10^{-11}$s for quantitative sample, we obtain $\Omega=0.63$GHz, which is of the same order of magnitude as typical period of laser-induced motion of magnetic bubble; GHz regime \cite{Moutafis}. Typical values of $\tau$ are in fact sub-picosecond\cite{WegrowePRB}\cite{OliveAPL}\cite{Fahnle}, which would push $\Omega$ beyond GHz regime according to Fig. 2. 
	
	\begin{figure}
		\includegraphics[angle=0,origin=c, scale=0.25]{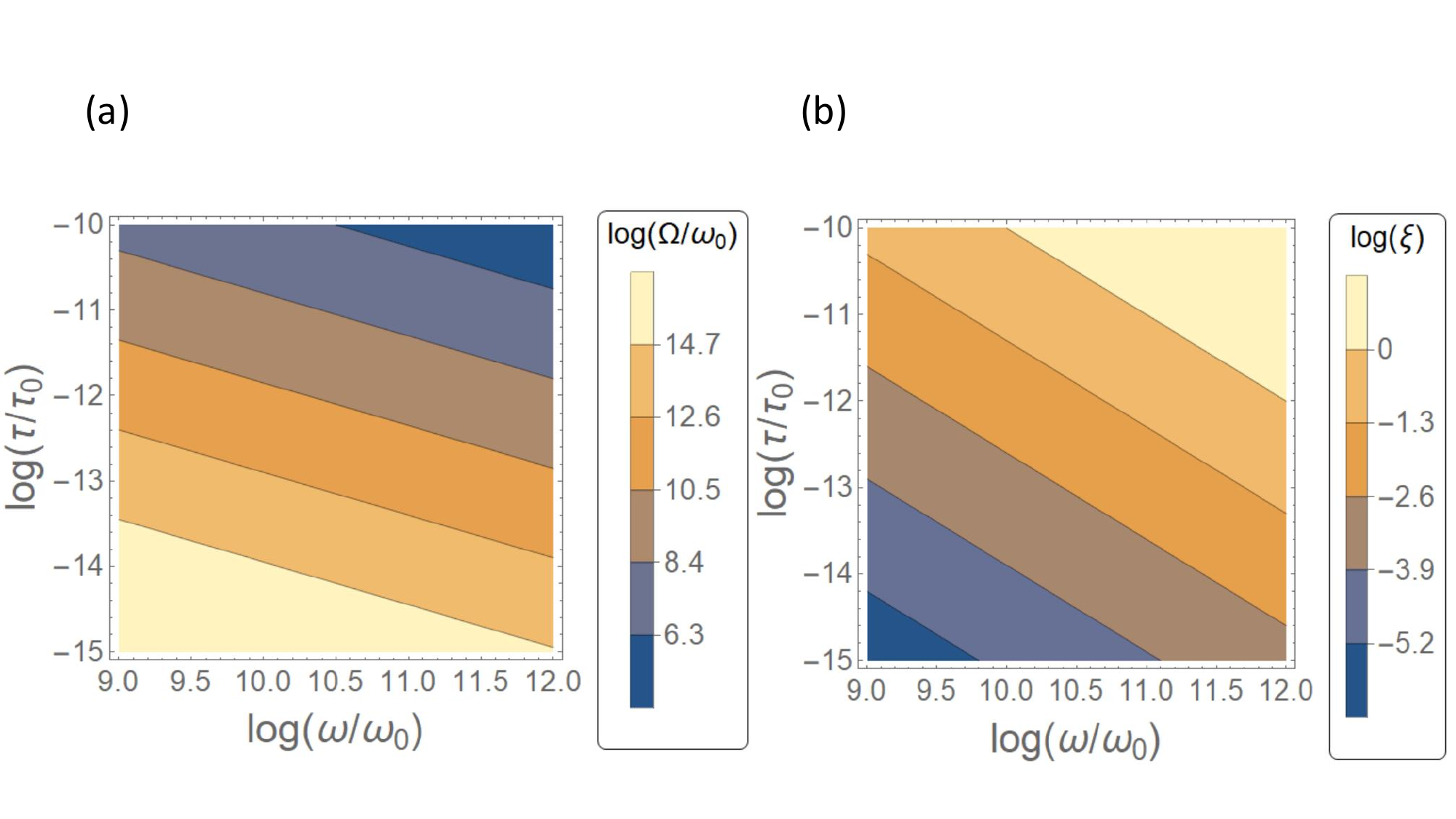}
		\caption{
			(a)The dependence of the Skyrmion cyclotron frequency $\Omega$(Hz) on the oscillator frequency $\omega$(Hz) and spin dynamics characteristic time $\tau$(s). Pertinent parameters are given in the text; the paragraph before Eq.(\ref{NetStaticForce}). The log is in base 10 and  reference quantities are $\Omega_0=\omega_0=1$Hz, $\tau_0=1$s.(b)The dependence of the perturbation parameter $\xi=\omega\tau$ and its validity regime $\omega\tau\leq 1$(or $\mathrm{log}(\omega\tau)\leq 0$). The resulting frequency $\Omega$ clearly enters ultrafast regime (beyond GHz (log$(\Omega/\Omega_0)\geq 9$)) for small $\omega\tau$.}\label{fig:Fig2}
	\end{figure}
	
	The net force
	\begin{equation}\label{NetStaticForce}
		\sum F_i=F_{2i}+F_{4i}+F_{Hi} 
	\end{equation}
	is static force and vanishes in equilibrium. This is obvious from physical principle that internal forces cannot drive a rigid body into motion. Mathematically speaking, $\sum \mathbf{F}=\nabla E_T=(\partial E_T/\partial\Theta)(\partial\Theta/\partial\mathbf{r})$. But $\partial E_T/\partial\Theta=0$ in equilibrium, when $\Theta$ takes the profile that minimizes $E_T$. As a result, $\sum \mathbf{F}=0$. When this force remains static but becomes nonzero for example in the presence of magnetic field gradient, the solution Eqs.(\ref{SolutionRx}-\ref{SolutionRy}) indicates that the Skyrmion drifts away from its initial position, while executing oscillatory (cyclotronic) motion. Our model Eq.(\ref{ModelHamiltonian}) does not contain an external conservative restoring force provided, for example, by a parabolic pinning potential $U(\mathbf{R})=K\mathbf{R}^2/2$ where $K$ is an effective spring constant, resulting in a spring force $\mathbf{F}=-K\mathbf{R}$. Adding this force to the right hand side of Eq.(\ref{EOMforR}) would lead to superposition of two cyclotronic motions of different frequencies \cite{MakhfudzPRL}. With this pinning potential,  a massless Skyrmion would also exhibit cyclotronic motion at a unique frequency $\Omega_{\mathrm{massless}}=K/(8\pi N_{Sk}\rho_s)$, with opposite dependence on the Skyrmion number $N_{Sk}$ and spin density $\rho_s$ to that of our massive Skyrmion Eq.(\ref{SkyrmionFrequency}). This difference in the dependence of cyclotron frequency on the Skyrmion number, that characterizes Skyrmion topology, can thus be used to distinguish our predicted massive Skyrmion dynamics, which can enter inertial regime, from massless Skyrmion dynamics, which may stay in adiabatic regime.  
	
	\textit{Discussion and Conclusion.\textemdash}A possible candidate material for which our study may be applicable is the NiGa$_2$S$_4$, where competing interactions have been proposed to host Skyrmion state\cite{MostovoyNatComm} and magnetoelastic coupling to phonon seems to be important \cite{BroholmPRL}. \textcolor{black}{Experimentally\cite{BroholmScience}\cite{BroholmPRLNiGa2S4}, it has been well understood that this material realizes an effective two-dimensional triangular lattice with ferromagnetic nearest-neighbor $J_1$ and antiferromagnetic third neighbor spin exchange interactions $J_3$, which are therefore frustrated and (in the presence of Zeeman field) reduce to Eq.(\ref{HSk}) in the continuum limit. Other candidate materials from frustrated magnets exist, such as $\alpha$-NaFeO$_2$ and Fe$_x$Ni$_{1-x}$Br$_2$, but NiGa$_2$S$_4$ is more promising due to its large $|J_3/J_1|$ ratio necessary to stabilize the Skyrmion, in addition to its well-studied magnetoelastic properties. Existing theory predicts that a single Skyrmion solution should be realizable in this material under the application of Zeeman field\cite{LinHayamiPRB}. Albeit a single Skyrmion solution can also emerge from unfrustrated spin-exchange interaction, Zeeman field, plus Dzyaloshinskii-Moriya interaction in chiral magnets, the Skyrmion size is always finite and tends to be small (nm scale) in this later class of magnets, rendering dissipative effect prominent. The large size of Skyrmion in inversion-symmetric magnets makes the latter the most suitable platform to test the prediction of our theory.} Our work also demonstrates that coupling to bosonic environment not only delivers a mechanism that generates a mass for the Skyrmion, but also promotes the dynamics of Skyrmion to inertial regime. The dynamics becomes fully deterministic in the classical limit of Skyrmion motion and can be viewed as arising from deterministic dynamics of the individual spins, which in inertial regime is described by the inertial Landau-Lifshitz-Gilbert equation. Thanks to the deterministic dynamics, we obtain a clear cut and simple expression for the Skyrmion mass, written in terms of intrinsic parameters of the model. This deterministic local-in-time dynamics can be traced to the ``natural coupling'' of the Skyrmion to the harmonic oscillators.
	
	While a massless Skyrmion may also exhibit a cyclotronic motion in the presence of a parabolic pinning potential, our massive Skyrmion exhibits a cyclotronic motion even in the absence of pinning potential, with a characteristic frequency that increases with Skyrmion number $N_{Sk}$, an opposite dependence on $N_{Sk}$(a topological invariant) to that of massless Skyrmion, a distinction that may serve as a smoking gun of the inertial Skyrmion dynamics predicted in this work. Our theory thus sheds light on the interplay of topology in magnetism and ultrafast spin dynamics. The deterministic nature of the equation of motion also suggests that controlling the frequency of the harmonic oscillators of the bosonic environment enables generation of tractable ultrafast dynamics for the Skyrmion. Concretely, this can be implemented by first generating and controlling the so-called coherent phonons, well studied in diverse areas experimentally\cite{ChoPRL,PfeiferPRL,AlbrechtPRL,YeePRL} and theoretically\cite{DresselhausPRB,MakhfudzPRApplied}, which then drives the Skyrmion motion via magnetoelastic coupling. Although our derivation has been done from a simplified model with a constant frequency spectrum for the boson, applicable to optical phonon, our theory can in principle be extended to proper phonon Hamiltonian involving both acoustic and optical phonons, \textcolor{black}{with a schematic outline for its derivation given in \cite{SupplementaryMaterials}}. The dispersive spectrum of the acoustic phonon enables variation of the phonon frequency, which thus allows the coherent control of the Skyrmion dynamics. Our proposal thus paves an avenue toward controllable high-frequency information storage technology based on Skyrmion and topological spin textures in general. 
	
	\textit{Acknowledgements.\textemdash}I.M. thanks M. Diouf, N. and R. Saidi, M. Ngom, C. Ngbo-Tiba Guiguisia, and M. Beye for the fruitful discussions during their internship with the author which inspired the latter to complete this work.

	\newpage
	
	\newpage
	
	\begin{widetext}
		
		\begin{center}
			\textbf{Supplementary Materials: Controllable and Non-Dissipative Inertial Dynamics of Skyrmion in a Bosonic Platform}\\
			\centering
			Imam Makhfudz
			
				\centering
			IM2NP, UMR CNRS 7334, Aix-Marseille Universit\'{e}, 13013 Marseille, France\\

		\date{\today}
		\bigskip
		
		\textcolor{black}{In these Supplementary Materials, we provide additional details supplementing the main text. First, we provide the derivation of effective action for the Skyrmion. Then, the derivation of the equation of motion from the effective action. Next, the link between the inertial spin dynamics to the Skyrmion equation of motion is discussed. The last two sections give a comparison between our minimal coupling and the more realistic magnetoelastic coupling and a comparison between our simple harmonic oscillator lattice model and the phonon Hamiltonian with both optical and acoustic phonons.} 
		
	\end{center}
	
	\section{Derivation of Skyrmion-Bath Action}
	We start from the bath Hamiltonian in the absence of Skyrmion,
	\begin{equation}\label{BathHamiltonianS}
		H_B=\hbar\omega\sum_{\alpha=x,y} \left(b^{\dag}_{\alpha}b_{\alpha}+\frac{1}{2}\right)
	\end{equation}
	where the operator $b^{\dag}_{\alpha},b_{\alpha}$ are dependent on the position $\mathbf{r}$; $b^{\dag}_{\alpha}=b^{\dag}_{\alpha}(\mathbf{r},t),b_{\alpha}=b_{\alpha}(\mathbf{r},t)$. In the presence of Skyrmion, adiabatically coupled to the bath degree of freedom, we have 
	\begin{equation}\label{BathSkyrmionHamiltonianS}
		H_{B+B-Sk}=\hbar\omega\sum_{\alpha=x,y} \left(b^{\dag}_{\alpha}(\mathbf{r}-\mathbf{R}(t),t)b_{\alpha}(\mathbf{r}-\mathbf{R}(t),t)+\frac{1}{2}\right)
	\end{equation}
	where $\mathbf{R}(t)$ is the center of mass position vector of the Skyrmion, serving as the collective coordinates. For a Skyrmion confined in a disk, it is natural to expect that its displacement would be small compared to both the Skyrmion radius $R_{Sk}$, the latter of which is of the same order of magnitude but smaller than the disk radius. We can thus expand the boson operators and retain liner order terms;
	\begin{equation}\label{BosonCreationOperatorShift}
		b^{\dag}_{\alpha}(\mathbf{r}-\mathbf{R}(t),t)=b^{\dag}_{\alpha}(\mathbf{r},t)-\mathbf{R}\cdot\frac{\partial b^{\dag}(\mathbf{r},t)}{\partial\mathbf{r}}
	\end{equation}
	\begin{equation}\label{BosonAnnihilationOperatorShift}
		b(\mathbf{r}-\mathbf{R}(t),t)=b_{\alpha}(\mathbf{r},t)-\mathbf{R}\cdot\frac{\partial b(\mathbf{r},t)}{\partial\mathbf{r}}.
	\end{equation}
	Substituting Eqs.(\ref{BosonCreationOperatorShift}-\ref{BosonAnnihilationOperatorShift}) into Eq.(\ref{BathSkyrmionHamiltonianS}), we obtain
	\[
	H_{B+B-Sk}=\hbar\omega\sum_{\alpha=x,y} \left(b^{\dag}_{\alpha}b_{\alpha}+\frac{1}{2}\right)
	\]
	\begin{equation}\label{HBBSklinear}
		-\hbar\omega\sum_{\alpha=x,y}R_i\left(b^{\dag}_{\alpha}(\mathbf{r},t)\frac{\partial b_{\alpha}(\mathbf{r},t)}{\partial r_i}.+ \frac{\partial b^{\dag}_{\alpha}(\mathbf{r},t)}{\partial r_i}b_{\alpha}(\mathbf{r},t)\right)
	\end{equation}
	where the last line describes the coupling between the Skyrmion collective coordinate and the bath operators, appearing in bilinear form, but which involves their spatial derivatives. 
	
	Next, we employ this Hamiltonian into the the time evolution operator, to derive the bath action and its coupling to the Skyrmion. At this point, the Hamiltonian is treated exactly, with no approximation. The time evolution operator is defined by \cite{KamenevBookS}
	\begin{equation}\label{TimeEvolutionOperator}
		\hat{\mathcal{U}}(t,t')=\mathbb{T}\exp\left(-i\int^t_{t'}H(t)dt\right)
	\end{equation}
	where $\mathbb{T}$ is time-ordering operator. In Keldysh theory, we would like to evaluate the partition function, given by
	\begin{equation}\label{PartitionFunction}
		Z=\frac{\mathrm{Tr[\mathcal{U}_C\rho]}}{\mathrm{Tr}[\rho]}
	\end{equation}
	where $\mathcal{U}_C$ is evaluated over the closed Keldysh contour (from $-\infty$ to $+\infty$ back to $-\infty$). The density matrix $\rho$ is taken to be equilibrium density matrix $\rho_0=\exp(-\beta(H-\mu N))$ where $\beta=1/(k_BT)$ is the inverse temperature, while $\mu$ is the chemical potential.
	
	We evaluate the numerator in Eq.(\ref{TimeEvolutionOperator}) using the discretization of the time integral in Eq.(\ref{TimeEvolutionOperator}), giving
	\begin{equation}\label{PartitionFunctionDiscrete}
		Z=\frac{1}{\mathrm{Tr}[\rho_0]}\int\Pi^{2N}_{j=1}d[\overline{\phi}_j,\phi_j]e^{-|\phi_j|^2}\langle \overline{\phi}_j|e^{-i\delta tH(\overline{\phi}_j,\phi_j)}|\phi_{j-1}\rangle
	\end{equation}
	where the time interval from $-\infty$ to $+\infty$ back to $-\infty$ has been discretized into $2N$ points ($N$ points for the forward path and $N$ points for the backward path). Our task is to evaluate the matrix element appearing in the integrand of the path integral, where $H$ is given by the $H_{B+B-Sk}$ in Eq.(\ref{HBBSklinear}). Tedious but straightforward calculation is detailed as follows.
	
	The matrix element of interest is given by
	\begin{equation}
		\langle \overline{\phi}_j|e^{-i\delta tH(\overline{\phi}_j,\phi_j)}|\phi_{j-1}\rangle \approx \langle \overline{\phi}_j|\left(1-i\delta tH(\overline{\phi}_j,\phi_j)\right)|\phi_{j-1}\rangle
	\end{equation}
	where the Hamiltonian to be used for $H(\overline{\phi}_j,\phi_j)$ is $H_{B+B-Sk}$. We therefore have to evaluate the matrix element
	\[
	\langle \overline{\phi}_j|H_{B+B-Sk}|\phi_j\rangle =  \langle \overline{\phi}_j|\hbar\omega\sum_{\alpha=x,y} \left(b^{\dag}_{\alpha}b_{\alpha}+\frac{1}{2}\right)-\hbar\omega\sum_{\alpha=x,y}R_i\left(b^{\dag}_{\alpha}(\mathbf{r},t)\frac{\partial b_{\alpha}(\mathbf{r},t)}{\partial r_i}.+ \frac{\partial b^{\dag}_{\alpha}(\mathbf{r},t)}{\partial r_i}b_{\alpha}(\mathbf{r},t)\right)|\phi_j\rangle
	\]
	\begin{equation}
		=  \langle \overline{\phi}_j|\hbar\omega\sum_{\alpha=x,y} \left(\overline{\phi}_{j\alpha}\phi_{j-1,\alpha}+\frac{1}{2}\right)-\hbar\omega\sum_{\alpha=x,y}\left(R^{j-1}_i\overline{\phi}_{j\alpha}(\mathbf{r},t)\frac{\partial \phi_{j-1,\alpha}(\mathbf{r},t)}{\partial r_i}+R^j_i \frac{\partial \overline{\phi}_{j\alpha}(\mathbf{r},t)}{\partial r_i}\phi_{j-1,\alpha}(\mathbf{r},t)\right)|\phi_j\rangle
	\end{equation}
	where we have used $\langle \overline{\phi}_j|b^{\dag}_{\alpha}=\langle \overline{\phi}_j|\overline{\phi}_{j,\alpha},b_{\alpha}|\phi_{j-1}\rangle=\phi_{j-1,\alpha}|\phi_{j-1}\rangle$. The superscript (subscript) in $R^j_i$ denotes the time discretization index (coordinate $i=x,y$) respectively. It is to be noted that the time index $j,j-1$ in the $R^j_i,R^{j-1}_i$ corresponds to the time index of the respective coherent state field $\overline{\phi}_{j\alpha},\phi_{j-1,\alpha}$ appearing within the partial derivative $\partial/\partial r_i$. Next, we express the variables with time discretization index $j-1$ as follows,
	\begin{equation}
		\phi_{j-1,\alpha}=\phi_{j,\alpha}-\delta t\frac{\partial \phi_{j,\alpha}}{\partial t},
		R^{j-1}_i=R^j_i-\tau\frac{\partial R^j_i}{\partial t}
	\end{equation}
	where $\tau$ is a new time scale, characteristic of the Skyrmion dynamics, which is distinct from the discretization time interval $\delta t$ characterizing the bath harmonic oscillator dynamics.

	Grouping the terms of similar forms and taking the continuum limit, $N\rightarrow \infty$, the final result gives
	\begin{equation}
		Z=\int D[\overline{\phi},\phi]e^{iS[\overline{\phi},\phi,\mathbf{R},\dot{\mathbf{R}}]}
	\end{equation}
	where $ D[\overline{\phi},\phi]=\Pi^{2N}_{j=1}d[\overline{\phi}_j,\phi_j]/\mathrm{Tr}[\rho_0]$ and the action is given by
	\begin{equation}
		S[\overline{\phi},\phi,\mathbf{R},\dot{\mathbf{R}}]=-\int dt \hbar\omega\sum_{\alpha}\frac{1}{2}+\int dt\sum_{\alpha}\overline{\phi}_{\alpha}(t)\left(i\partial_t-\hbar\omega\right)\phi_{\alpha}(t)-\int dt\sum_{\alpha}\overline{\phi}_{\alpha}(t)\left(\hbar\omega\tau\dot{\mathbf{R}}\cdot i\nabla\right)\phi_{\alpha}(t)+\int dt\hbar\omega\sum_{\alpha}\mathbf{R}\cdot\nabla|\phi_{\alpha}(t)|^2
	\end{equation}    
	where we have introduced a time scale constant $\tau$, that describes the characteristic time scale of the dynamics of the Skyrmion, which is distinct from that of the heat bath harmonic oscillator. The first term is a constant action term that will be dropped from further consideration. There are two terms that reflect the coupling of the Skyrmion to the heat bath degree of freedom, one is proportional o the Skyrmion velocity $\dot{\mathbf{R}}$, the other is proportional to the Skyrmion position vector $\mathbf{R}$. Both terms carry effective coupling constant proportional to the frequency $\omega$ of the harmonic oscillator of the heat bath. When deriving the effective action for the Skyrmion due to its coupling to the heat bath, one can therefore define a perturbation parameter proportional to the ratio between the harmonic oscillator energy and the Skyrmion magnetic energy.
	
	Evaluating the action around closed time contour of Keldysh (from $t=-\infty$ to $t=+\infty$ and back to $t=-\infty$) \cite{KamenevBookS}, we arrive at the action for the bath,
	\begin{equation}
		S_{B+B-Sk}=\int^{+\infty}_{-\infty}dt\int d\mathbf{r} \sum_{\alpha=x,y}\left[\overline{\phi}^+_{0\alpha}(\mathbf{r},t)G^{-1}_{\mathbf{R}^+}(\omega,t)\phi^+_{0\alpha}(\mathbf{r},t)-\overline{\phi}^-_{0\alpha}(\mathbf{r},t)G^{-1}_{\mathbf{R}^-}(\omega,t)\phi^-_{0\alpha}(\mathbf{r},t)\right]
	\end{equation}
	where the overall minus sign in the second term comes from compensating the switching the limits of time integral. The $\phi^{+(-)}_0(\mathbf{r},t)$ represent the forward (backward) coherent state field of the bath oscillator. Similarly with $\mathbf{R}^{+(-)}$ for the collective coordinates of the Skyrmion. We then rotate the variables into their classical and quantum counterparts, 
	\begin{equation}
		\phi^{\pm}_0=\frac{1}{\sqrt{2}}\left(\phi^{cl}_0\pm\phi^q_0\right)
	\end{equation}
	\begin{equation}
		\mathbf{R}^{\pm}=\mathbf{R}_c\pm\frac{\mathbf{R}_q}{2}
	\end{equation}
	where the superscript or subscript $cl,q$ represent the classical and quantum parts, respectively.  
	The resulting double-time integral action for the bath field and its coupling to Skyrmion is
	\begin{equation}
		S_{B+B-Sk}= S_{B}+ S_{B-Sk}
	\end{equation}
	where the non-interacting bath harmonic oscillator action is given by
	\begin{equation}
		S_{B}=\int^{+\infty}_{-\infty}dt \int^{+\infty}_{-\infty}dt'\int d\mathbf{r}\sum_{\alpha=x,y} \left[(\overline{\phi}^{cl}_{0\alpha}(\mathbf{r},t),\overline{\phi}^{q}_{0\alpha}(\mathbf{r},t))
		\begin{pmatrix}
			0 & [G^{-1}]^A \\
			[G^{-1}]^R & [G^{-1}]^K 
		\end{pmatrix}_{t,t'}	
		\begin{pmatrix} \phi^{cl}_{0\alpha} (\mathbf{r},t')\\ \phi^q_{0\alpha}(\mathbf{r},t') 
		\end{pmatrix}\right]
	\end{equation}
	where the Green functions are given by
	\begin{equation}
		[G^{-1}]^A(t,t') = \left(i\partial_t-\omega-i0\right)\delta(t-t')
	\end{equation}
	\begin{equation}
		[G^{-1}]^R(t,t') = \left(i\partial_t-\omega+i0\right)\delta(t-t')
	\end{equation}
	\begin{equation}
		[G^{-1}]^K (t,t')=2i0 F(t,t')
	\end{equation}
	where $F(t,t')$ is a Hermitian matrix. This last Green function (Keldysh) is thus infinitesimally small, which is a characteristic of noninteracting function. The space-time expression for the above Green functions are given by
	\begin{equation}
		G^R (t,t')=-i\theta(t-t')e^{-i\omega(t-t')}
	\end{equation}
	\begin{equation}
		G^A (t,t')=i\theta(t'-t)e^{-i\omega(t-t')}.
	\end{equation}
	
	The action for the bath-Skyrmion coupling is 
	\begin{equation}
		S_{B-Sk}=      
		\int^{+\infty}_{-\infty}dt \int^{+\infty}_{-\infty}dt'\int d\mathbf{r}\sum_{\alpha=x,y} \left[(\overline{\phi}^{cl}_{0\alpha}(\mathbf{r},t),\overline{\phi}^{q}_{0\alpha}(\mathbf{r},t))
		\begin{pmatrix}
			[\tilde{G}^{-1}]^{cl,cl} & [\tilde{G}^{-1}]^{cl,q} \\
			[\tilde{G}^{-1}]^{q,cl} & [\tilde{G}^{-1}]^{q,q} 
		\end{pmatrix}_{t,t'}	
		\begin{pmatrix} \phi^{cl}_{0\alpha} (\mathbf{r},t')\\ \phi^q_{0\alpha}(\mathbf{r},t') 
		\end{pmatrix}\right]
	\end{equation}
	where the Green functions are given by
	
	\begin{equation}
		[\tilde{G}^{-1}]^{cl,cl}(t,t') = -\frac{i}{2}\dot{\mathbf{R}}_q\cdot\frac{\partial}{\partial \mathbf{r}}\delta(t-t')
	\end{equation}
	\begin{equation}
		[\tilde{G}^{-1}]^{cl,q}(t,t') =  -i\dot{\mathbf{R}}_c\cdot\frac{\partial}{\partial \mathbf{r}}\delta(t-t')
	\end{equation}
	\begin{equation}
		[\tilde{G}^{-1}]^{q,cl}(t,t') = -i\dot{\mathbf{R}}_c\cdot\frac{\partial}{\partial \mathbf{r}}\delta(t-t')
	\end{equation}
	\begin{equation}
		[\tilde{G}^{-1}]^{q,q} (t,t')=-\frac{i}{2}\dot{\mathbf{R}}_q\cdot\frac{\partial}{\partial \mathbf{r}}\delta(t-t').
	\end{equation}
	Eventually, we will retain only zeroth order terms in $\dot{\mathbf{R}}_q$, giving Green functions that agree with those with the required structure to produce proper correlation function.

	\section{Derivation of Equation of Motion of Skyrmion Center-of-Mass Position}
	
	We start from the effective action for the Skyrmion upon integrating out the bath degree of freedom,
	\begin{equation}
		S^{eff}_{Sk}[\mathbf{R},\dot{\mathbf{R}}]=S_{Sk}[\mathbf{R},\dot{\mathbf{R}}]+\delta S[\mathbf{R},\dot{\mathbf{R}}]
	\end{equation}
	where 
	\[
	S_{Sk}[\mathbf{R},\dot{\mathbf{R}}]= 4\pi N_{Sk}\rho_s \int \left(R_x\dot{R}_y-R_y\dot{R}_x\right)dt
	\]
	\begin{equation}
		-\rho_s\int dt\int d^2\mathbf{r}\dot{\mathbf{R}}\cdot\left(1-\cos\Theta_0\right)\nabla\Phi_0
		+\int dt \left(\mathbf{F}_2+\mathbf{F}_4+\mathbf{F}_H\right)\cdot \mathbf{R}
	\end{equation}
	and
	\begin{equation}\label{SkyrmionActionCorrectionS}
		\delta S[\mathbf{R},\dot{\mathbf{R}}]=\frac{(\hbar\omega\tau)^2}{16\pi}\Lambda^4 \int dt \left(\dot{R}^2_x+\dot{R}^2_y\right).
	\end{equation}
	The equations of motion for $R_x,R_y$ are derived using Euler-Lagrange equation,
	\begin{equation}
		\frac{\partial}{\partial t}\frac{\delta S}{\delta \dot{R}_i}-\frac{\delta S}{\delta R_i}=0
	\end{equation}
	where $i=x,y$. Applying the above equation to the effective action $S^{eff}_{Sk}$ gives the equations of motion Eq.(\ref{EOMforR}) in the main text. 
	
	\section{Link to Inertial Spin Dynamics}
	Our result demonstrates that the coupling of the Skyrmion to the harmonic oscillators bath induces a deterministic massive dynamics for the Skyrmion. It will be argued here that this deterministic massive dynamics arises from the deterministic inertial dynamics of the individual spins, of the thousands of spins constituting the Skyrmion. Recent work  \cite{DuinePRLinertiaS} demonstrates that linearly coupling a spin (or macrospin) to harmonic oscillators bath induces an inertia to the spin. The spin obeys the following equation of motion upon integrating out the harmonic oscillators \cite{DuinePRLinertiaS},
	\begin{equation} \dot{\mathbf{S}}=\mathbf{S}\times \left(\mathbf{B}+\zeta\right)+\mathbf{S}\times\int^{\infty}_{-\infty}dt'\tilde{\alpha}(t-t')\mathbf{S}(t')
	\end{equation}
	where $\zeta$ is a fluctuating field with a white noise Gaussian spectrum; $\langle \zeta_m(t)\rangle=0$ and $\langle \zeta_m(t)\zeta_{m'}(t')\rangle=-(i/2\hbar)\delta_{mm'}\alpha^K(t-t')$. Now, consider the local-in-time limit of the above equation, obtained by writing 
	\begin{equation}
		\tilde{\alpha}(t-t')=\delta(t-t')\left(\tilde{\alpha}_0+\tilde{\alpha}_1\frac{\partial}{\partial t}+\tilde{\alpha}_2\frac{\partial^2}{\partial t^2}\right),
	\end{equation}
	we then immediately obtain the by now well-established local-in-time form of the ``inertial Landau-Lifshitz-Gilbert equation''\cite{WegrowePRBs},
	\begin{equation}\label{iLLG} \dot{\mathbf{S}}=\mathbf{S}\times \mathbf{B}-\alpha\mathbf{S}\times\frac{\partial \mathbf{S}}{\partial t}-\alpha\tau\mathbf{S}\times\frac{\partial^2\mathbf{S}}{\partial t^2}
	\end{equation}
	where we have absorbed the fluctuating field $\zeta$ into the effective magnetic field $\mathbf{B}$, identified $\tilde{\alpha}_1$ with the standard Gilbert damping constant $\alpha$ and written $\tilde{\alpha}_2=-\alpha\tau$, where $\tau$ is the inertial time appearing earlier in the Skyrmion dynamics. Omitting the damping $\alpha$ term, one can directly see that this spin equation of motion takes analogous form as our equation of motion for Skyrmion Eq.(25) in the main text, rewritten here for convenience 
	\begin{equation}\label{EOMforRs}
		A\frac{\hbar\omega\tau^2}{8\pi}\Lambda^4\ddot{R}_i+8\eta_i\pi N_{Sk}\rho_s\dot{R}_{\tilde{i}}=F_{2i}+F_{4i}+F_{Hi}
	\end{equation}
	where $i=x,y$, when $\tilde{i}=y,x$ and $\eta_i=-1(+1)$ for $i=x(y)$
	where the mass term (the first term on left hand side) in Eq.(\ref{EOMforRs}) corresponds to the inertial term in Eq.(\ref{iLLG}), while for a spin or macrospin, the topological term does not exist, as it corresponds to $N_{Sk}=0$, representing a uniform ferromagnet with parallel magnetization as the direct analog of a macrospin. The implication is clear; first, the inertia of the individual spins, induced by their coupling to the harmonic oscillators bath, generates inertia (mass) mass to Skyrmion. Second, the deterministic form of the by-now-established inertial Landau-Lifshitz-Gilbert equation in fact drives deterministic dynamics for the Skyrmion, as confirmed by our analytical result, represented by the Skyrmion equation of motion Eq.(\ref{EOMforRs}).  
	
	\section{Comparison to Magnetoelastic Coupling and Its Effect on Skyrmion Dynamics}
	
	In the model presented in the main text, we have proposed a ``natural but minimal coupling'' to describe the simplest form of the coupling between the Skyrmion and the phononic oscillators. One may ask how to justify such a coupling, what its coupling parameters are, and how it relates to the more familiar magnetoelastic coupling usually used in the literature to model the interaction between the magnetic and lattice degrees of freedom. We will take a concrete example of magnetoelastic coupling described in \cite{RuckriegelPRBsupp} to make quantitative analysis and connection to our minimal coupling. Such magnetoelastic coupling is linear in the phonon displacement field $\mathbf{u}$, in terms of the strain tensor $u_{\alpha\beta}=\left(\partial_{\beta} u_{\alpha}+\partial_{\alpha}u_{\beta}\right)/2$ which for the exchange part takes the form $B'_{\alpha\beta}\partial_{\alpha}\mathbf{S}\cdot\partial_{\beta}\mathbf{S}u_{\alpha\beta}$, where $B'_{\alpha\beta}$ is the magnetoelastic coupling constant and $\partial_{\alpha(\beta)}=\partial/\partial r_{\alpha(\beta)}$\cite{RuckriegelPRBsupp}. Upon considering a Skyrmion, the magnetoelastic coupling turns into $-B'_{\alpha\beta}\mathbf{R}\cdot\nabla\left(\partial_{\alpha}\mathbf{S}\cdot\partial_{\beta}\mathbf{S}\left(\partial_{\beta} u_{\alpha}+\partial_{\alpha}u_{\beta}\right)\right)/2$, which therefore involves in overall  first and second derivatives of the phonon displacement field corresponding to respectively terms dependent linearly and quadratically on the phonon wave vector $\mathbf{k}$, compared to the pure linear $\mathbf{k}$ dependence in our minimal coupling. While our minimal coupling is a higher order term in the phonon displacement field $\mathbf{u}$, such a quadratic term is allowed by symmetry \cite{KaganovTsukerniksupp} and may eventually dominate over the magnetoelastic coupling when the phonons are excited at resonance; i.e. at the frequency of the flat band corresponding to the optical phonon-like mode frequency, due to the extremely large phonon density of states there. One can also say that our minimal coupling will prevail in the long distance limit, corresponding to small wave vector $k=|\mathbf{k}|$, because the minimal coupling is of lower order in $k$ than the magnetoelastic coupling.
	
	In the following, we derive the resulting dynamics of Skyrmion in the presence of (acoustic) phonon and magnetoelastic coupling to the phonon. The magnetoelastic coupling in magnetic materials consists of two contributions; the one coming from magnetostatic (dipolar) interaction, dominant in long distance regime, and the one arising from magnetic exchange interaction, dominant in the short distance regime. The corresponding Hamiltonian is given by (Eq.(3.1) of \cite{RuckriegelPRBsupp})
	\begin{equation}
		E_{me}\left[\mathbf{M},\mathbf{u}\right]=\frac{n d}{M^2_s}\int d^2r
		\sum_{\alpha\beta}\left(B_{\alpha\beta}M_{\alpha}(\mathbf{r})M_{\beta}(\mathbf{r})u_{\alpha\beta}(\mathbf{r})+B'_{\alpha\beta}\frac{\partial \mathbf{M}(\mathbf{r})}{\partial r_{\alpha}}\cdot\frac{\partial \mathbf{M}(\mathbf{r})}{\partial r_{\beta}}u_{\alpha\beta}(\mathbf{r})\right)
	\end{equation}
	where $n$ is the number density of magnetic ions, $M_s$ the saturation magnetization, $B_{\alpha\beta},B'_{\alpha\beta}$ the magnetoelastic couplings for the dipolar and exchange parts respectively, $M_{\alpha}$ the component of magnetization along direction $\alpha=x,y,z$ while the strain tensor is given by
	\begin{equation}\label{straintensor}
		u_{\alpha\beta}=\frac{1}{2}\left[\frac{\partial u_{\alpha}(\mathbf{r})}{\partial r_{\beta}}+\frac{\partial u_{\beta}(\mathbf{r})}{\partial r_{\alpha}}\right],
	\end{equation}
	where we have substituted $\mathbf{u}$ for $\mathbf{X}$ to conform with our notation in the main text, representing the phonon displacement by $\mathbf{u}$ instead of $\mathbf{X}$ used in \cite{RuckriegelPRBsupp}. We have also replaced the 3D integral $\int d^3\mathbf{r}=d\int d^2\mathbf{r}$ where $d$ is the thickness of the sample, implying that the physics is independent of $z$ and the system effectively behaves as a 2D system valid for small enough thickness $d$ along $z$. In the long wave length limit, it is justified to retain only the dipolar interaction, while exchange interaction is neglected, as done in \cite{RuckriegelPRB}. In our study, the Skyrmion is stabilized by exchange spin interactions. Therefore, for this purpose of comparing magnetoelastic coupling to our minimal coupling, we will retain the exchange part and omit the dipolar part in the magnetoelastic coupling.

	In the presence of Skyrmion, and within rigid Skyrmion picture, the magnetoelastic energy is transformed into a shifted form obtained by rewriting $M_{\alpha}(\mathbf{r})\rightarrow M_{\alpha}(\mathbf{r}-\mathbf{R}),u_{\alpha\beta}(\mathbf{r})\rightarrow u_{\alpha\beta}(\mathbf{r}-\mathbf{R})$. Expanding in $\mathbf{R}$, the linear-in-$\mathbf{R}$ order term is given by
	\begin{equation}\label{linearMEenergy}
		E^{linear-\mathbf{R}}_{me}
		=-\frac{n d}{M^2_s}\int d^2\mathbf{r}\sum_{\alpha\beta}B'_{\alpha\beta}\mathbf{R}\cdot
		\nabla\left[\partial_{\alpha}\mathbf{M}(\mathbf{r})\cdot\partial_{\beta}\mathbf{M}(\mathbf{r})u_{\alpha\beta}(\mathbf{r})\right]
	\end{equation}
	which is linear in the phonon displacement vector field $\mathbf{u}(\mathbf{r})$ via the strain tensor Eq.(\ref{straintensor}). Now, let us compare this magnetoelastic energy coupling the magnetization and phonon to the uncoupled harmonic oscillators' energy given in the main text. From the displacement part of the harmonic oscillator energy given in Eq.(3) in the main text, we have
	\begin{equation}
		E_{phonon}=d\int d^2\mathbf{r}\left(\frac{1}{2}\rho\dot{\mathbf{u}}^2+\frac{1}{2}\rho\omega^2\mathbf{u}^2\right)
	\end{equation}
	where we have used $\int d^3\mathbf{r}$ like before. In the presence of Skyrmion, we shift $u(\mathbf{r})\rightarrow u(\mathbf{r}-\mathbf{R})$ and expand in $\mathbf{R}$. The linear in $\mathbf{R}$ order term is given by
	\begin{equation}\label{minimalcouplingenergy}
		E^{linear-\mathbf{R}}_{phonon}=- d\int d^2\mathbf{r}\mathbf{R}\cdot \nabla\left(\frac{1}{2}\rho\dot{\mathbf{u}}^2+\frac{1}{2}\rho\omega^2\mathbf{u}^2\right)
	\end{equation}
	which can be compared directly with the magnetoelastic coupling energy Eq.(\ref{linearMEenergy}). One observes that while magnetoelastic coupling is linear in the phonon displacement field $\mathbf{u}$, the minimal but natural coupling adopted in the present work is quadratic in $\mathbf{u}$.    
	
	Now, let us consider a model with such magnetoelastic coupling rather than our minimal coupling. To simplify the derivation, without loss of generality, let us assume diagonal and isotropic exchange magnetoelastic coupling, by taking $B'_{\alpha\beta}=B'_0\delta_{\alpha\beta}$. This magnetoelastic coupling accompanies the phonon Hamiltonian, which in real space is given by
	\begin{equation}
		\label{Hphonon}H_{phonon}=\sum_{\mathbf{r}}\frac{\mathbf{p}^2(\mathbf{r})}{2m}+\frac{1}{2}m\omega^2\sum_{\mathbf{r},\mathbf{r}'}(\mathbf{u}(\mathbf{r}')-\mathbf{u}(\mathbf{r}))^2
	\end{equation}
	where $\mathbf{p}(\mathbf{r})$ and $\mathbf{u}(\mathbf{r})$ are respectively the momentum and displacement of the ion at position $\mathbf{r}$ in the lattice. Again, we have assumed diagonal and isotropic potential energy for simplicity; $\omega_{ij}=\omega\delta_{ij}$. Next, we diagonalize the Hamiltonian from real space to reciprocal space, by defining
	\begin{equation}
		Q_{\mathbf{k}\alpha}=\frac{1}{\sqrt{N}}\sum_{\mathbf{r}}e^{i\mathbf{k}\cdot\mathbf{r}}u_{\alpha}(\mathbf{r}),\Pi_{\mathbf{k}\alpha}=\frac{1}{\sqrt{N}}\sum_{\mathbf{r}}e^{i\mathbf{k}\cdot\mathbf{r}}p_{\alpha}(\mathbf{r}),
	\end{equation}
	one arrives at
	\begin{equation}
		H_{phonon}=\frac{1}{2m}\sum_{\mathbf{k}\alpha}\left(\Pi_{\mathbf{k}\alpha}\Pi_{-\mathbf{k}\alpha}+m^2\omega^2_{\mathbf{k}}Q_{\mathbf{k}\alpha}Q_{-\mathbf{k}\alpha}\right)
	\end{equation}
	which is diagonal in the $\mathbf{k}$ space, with a dispersion relation for the phonon $\omega_{\mathbf{k}}$ which depends on the lattice detail, which we will keep in its general form without evaluating it explicitly. Next, we quantize the conjugate variables by writing
	\begin{equation}\label{quantization}
		Q_{\mathbf{k}}=\sqrt{\frac{\hbar}{2m\omega_{\mathbf{k}}}}\left(b^{\dagger}_{\mathbf{k}}+b_{-\mathbf{k}}\right),\Pi_{\mathbf{k}}=i\sqrt{\frac{\hbar m\omega_{\mathbf{k}}}{2}}\left(b^{\dagger}_{\mathbf{k}}-b_{-\mathbf{k}}\right)
	\end{equation}, giving quantum Hamiltonian of the phonon in the reciprocal space
	\begin{equation}
		H_{phonon}=\sum_{\mathbf{k}\alpha}\hbar\omega_{\mathbf{k}}\left(b^{\dagger}_{\mathbf{k}\alpha}b_{\mathbf{k}\alpha}+\frac{1}{2}\right)
	\end{equation}
	where the bosonic operators act on the coherent states as $b_{\mathbf{k}\alpha}|\phi_{\alpha}(\mathbf{k})\rangle=\phi_{\alpha}(\mathbf{k})|\phi_{\alpha}(\mathbf{k})\rangle, \langle\overline{\phi}_{\alpha}(\mathbf{k})|b^{\dagger}_{\mathbf{k}\alpha}=\langle\overline{\phi_{\alpha}}(\mathbf{k})|\overline{\phi}_{\alpha}(\mathbf{k})$. Such a simple monoatomic model given by Eq.(\ref{Hphonon}) would eventually give an acoustic phonon mode only; for one-dimensional atomic chain for example, we have $\omega_k=2\omega|\sin k/2|$, which is linear in $k$ at small $k$ and vanishes at $k=0$.

	Following the same path integral approach in Section I, one will arrive at the following effective action written in the $\mathbf{k}$ space,
	\begin{equation}\label{SphiRsupp0}
		S[\overline{\phi},\phi,\mathbf{R},\dot{\mathbf{R}}]=\int dt  \sum_{\mathbf{k}\alpha}\left[\overline{\phi}_{\alpha}(\mathbf{k},t)\hbar\left(i\partial_t-\omega_{\mathbf{k}}\right)\phi_{\alpha}(\mathbf{k},t)
		+\frac{inB'_0}{\sqrt{N}}k_{\alpha}\sqrt{\frac{\hbar}{2m\omega_{\mathbf{k}}}}\left(\overline{\phi}_{\alpha}(\mathbf{k},t)+\phi_{\alpha}(-\mathbf{k},t)\right)\sum_{\mathbf{r}}\left(\mathbf{R}\cdot\nabla-\tau\dot{\mathbf{R}}\cdot i\nabla\right)C_{\alpha\mathbf{k}}(\mathbf{r})\right]
	\end{equation}
	where
	\begin{equation}
		C_{\alpha\mathbf{k}}(\mathbf{r})=\sum_i\left(\frac{\partial n_i(\mathbf{r})}{\partial r_{\alpha}}\right)^2e^{-i\mathbf{k}\cdot\mathbf{r}}
	\end{equation}
	for which we have defined
	\begin{equation}
		n_i(\mathbf{r})=\frac{M_i(\mathbf{r})}{M_s}.
	\end{equation}
	Using the functional integral formula\cite{KamenevBookS}
	\begin{equation}\label{pathintegralformula}
		Z[\overline{J},J]=\int d[\overline{\phi},\phi]e^{-\overline{\phi}\hat{A}\phi+\overline{\phi}J+\overline{J}\phi}=\frac{e^{\overline{J}\hat{A}^{-1}J}}{\mathrm{det}\hat{A}}
	\end{equation}
	one can already see that the coupling part (the part with overall prefactor $inB'_0/\sqrt{N}$ within the square brackets in the action Eq.(\ref{SphiRsupp0})) will produce additional terms in the action for $\mathbf{R}$ that are of the form $\mathbf{R}^2, R_i\dot{R}_j$, and $\dot{\mathbf{R}}^2$. The first term behaves as behaves as a confining potential. This is quite intuitive; the magnetoelastic coupling pulls the Skyrmion and keep it bound to its equilibrium position. The second term acts like dissipative friction energy. The third term acts like a kinetic energy that gives a mass to the Skyrmion dynamics. The familiar magnetoelastic coupling can thus produce a mass for the Skyrmion. But on top of that, it also produces confining potential and dissipative friction force in the Skyrmion dynamics. In a sense, our minimal coupling is simpler than the familiar magnetoelastic coupling; while the minimal coupling produces mass term in the Skyrmion equation of motion and dissipationless dynamics, magnetoelastic coupling also produces confining potential energy and dissipative friction terms in addition to a mass term.
	
	More precisely, evaluating the path integral over the bosonic coherent state fields $\overline{\phi},\phi$ and working in Keldysh space like done in Section I, we obtain
	\begin{equation}
		Z[\mathbf{R},\dot{\mathbf{R}}]=\int d\left[\overline{\phi},\phi\right]e^{iS[\overline{\phi},\phi,\mathbf{R},\dot{\mathbf{R}}]}=e^{iS_{eff}[\mathbf{R},\dot{\mathbf{R}}]}
	\end{equation}
	where $S_{eff}[\mathbf{R},\dot{\mathbf{R}}]$ is the resulting effective action for the Skyrmion collective coordinates. Using the path integral formula Eq.(\ref{pathintegralformula}) with 
	\begin{equation}
		S[\overline{\phi},\phi,\mathbf{R},\dot{\mathbf{R}}]=S_{B}+\overline{\phi}J+\overline{J}\phi
	\end{equation}
	for which $\overline{\phi}=(\overline{\phi}^{cl}_{0\alpha}(\mathbf{k},t),\overline{\phi}^{q}_{0\alpha}(\mathbf{k},t))$ and
	\begin{equation}
		S_{B}=\int^{+\infty}_{-\infty}dt \int^{+\infty}_{-\infty}dt'\sum_{\mathbf{k}}\sum_{\alpha=x,y} \left[(\overline{\phi}^{cl}_{0\alpha}(\mathbf{k},t),\overline{\phi}^{q}_{0\alpha}(\mathbf{k},t))
		\begin{pmatrix}
			0 & [D^{-1}]^A \\
			[D^{-1}]^R & [D^{-1}]^K 
		\end{pmatrix}_{t,t'}	
		\begin{pmatrix} \phi^{cl}_{0\alpha} (\mathbf{k},t')\\ \phi^q_{0\alpha}(\mathbf{k},t') 
		\end{pmatrix}\right]
	\end{equation}
	\comment{where the Green functions are given by
		\begin{equation}
			[D^{-1}]^A(t,t') = \left(i\partial_t-\omega_{\mathbf{k}}-i0\right)\delta(t-t')
		\end{equation}
		\begin{equation}
			[D^{-1}]^R(t,t') = \left(i\partial_t-\omega_{\mathbf{k}}+i0\right)\delta(t-t')
		\end{equation}
		\begin{equation}
			[D^{-1}]^K (t,t')=2i0 F(t,t')
		\end{equation}
		where $F(t,t')$ is a Hermitian matrix,}
	while 
	\begin{equation}
		J=inB'_0\sqrt{\hbar/(2mN\omega_{\mathbf{k}})}
		\hat{\sigma}_1
		\sum_{\alpha}k_{\alpha}\sum_{\mathbf{r}}\left(\mathbf{R}\cdot\nabla-\tau\dot{\mathbf{R}}\cdot i\nabla\right)C_{\alpha\mathbf{k}}(\mathbf{r})
	\end{equation}
	where $\omega_{-\mathbf{k}}=\omega_{\mathbf{k}}$ has been assumed and $\hat{\sigma}_1$ is the $x$ Pauli matrix in the Keldysh space, one arrives at 
	\begin{equation}
		S_{eff}[\mathbf{R},\dot{\mathbf{R}}]=\frac{1}{2}\frac{n^2B^{'2}_0\hbar^2}{2mN}\int\int^{+\infty}_{-\infty}dtdt'\sum_{\mathbf{k}\alpha\alpha'}\frac{k_{\alpha}k_{\alpha'}}{\omega_{\mathbf{k}}}\sum_{\mathbf{r}\mathbf{r}'}\left(\mathbf{R}^T(t)\cdot\nabla-\tau\dot{\mathbf{R}}^T(t)\cdot i\nabla\right)C^*_{\alpha\mathbf{k}}\left(\mathbf{r}\right)\mathcal{D}^{-1}(t-t')\left(\mathbf{R}(t')\cdot\nabla-\tau\dot{\mathbf{R}}(t')\cdot i\nabla\right)C_{\alpha'\mathbf{k}}(\mathbf{r}')
	\end{equation}
	where $\mathbf{R}^T=(\mathbf{R}_{cl},\mathbf{R}_q)$ and $\dot{\mathbf{R}}^T=(\dot{\mathbf{R}}_{cl},\dot{\mathbf{R}}_q)$, while the Green's function matrix is given by
	\begin{equation}
		\mathcal{D}^{-1}(t-t')=-\hat{\sigma}_1\sum_{\alpha}D_{\alpha}(t-t')\hat{\sigma}_1
	\end{equation}
	for which 
	\begin{equation}
		D^{-1}_{\alpha}(t-t')=\begin{pmatrix}
			0 & [D^{-1}]^A\\
			[D^{-1}]^R & [D^{-1}]^K
		\end{pmatrix}
	\end{equation}
	with the Green's functions $[D^{-1}]^{R(A)}(t,t'), [D^{-1}]^K (t,t')$ which now depend on the spectrum of the bath. The determinant $\mathrm{det}\hat{A}$ in the denominator of Eq.(\ref{pathintegralformula}) can always be normalized to unity, corresponding to normalization of the partition function, by proper choice of integration measure \cite{KamenevBookS}.

	Before we proceed, we evaluate the following expression explicitly, 
	\begin{equation}
		\mathbf{R}\cdot\nabla C_{\alpha\mathbf{k}}(\mathbf{r})=e^{-i\mathbf{k}\cdot\mathbf{r}}\left(R_xf_{\alpha x}(k_x)+R_yf_{\alpha y}(k_y)\right), \dot{\mathbf{R}}\cdot\nabla C_{\alpha\mathbf{k}}(\mathbf{r})=e^{-i\mathbf{k}\cdot\mathbf{r}}\left(\dot{R}_xf_{\alpha x}(k_x)+\dot{R}_yf_{\alpha y}(k_y)\right)
	\end{equation}
	where
	\begin{equation}
		f_{\alpha x}(k_x)=\sum_i\left[2\left(\frac{\partial n_i(\mathbf{r})}{\partial r_{\alpha}}\right)\frac{\partial^2n_i(\mathbf{r})}{\partial x\partial r_{\alpha}}-ik_x\left(\frac{\partial n_i(\mathbf{r})}{\partial r_{\alpha}}\right)^2\right], f_{\alpha y}(k_y)=\sum_i\left[2\left(\frac{\partial n_i(\mathbf{r})}{\partial r_{\alpha}}\right)\frac{\partial^2n_i(\mathbf{r})}{\partial y\partial r_{\alpha}}-ik_y\left(\frac{\partial n_i(\mathbf{r})}{\partial r_{\alpha}}\right)^2\right]
	\end{equation}

	Expanding the action, we can write 
	\begin{equation}
		S_{eff}[\mathbf{R},\dot{\mathbf{R}}]=S_{pot}[\mathbf{R}]+S_{kin}[\dot{\mathbf{R}}]+S_{diss}[\mathbf{R},\dot{\mathbf{R}}]
	\end{equation}
	where 
	\[
	S_{pot}[\mathbf{R}]= - \frac{1}{2}\frac{n^2B^{'2}_0\hbar^2}{2mN}\int dt\int^{+\infty}_{-\infty}dt'\sum_{\mathbf{k}}\sum_{\alpha\alpha'}\frac{k_{\alpha}k_{\alpha'}}{\omega_{\mathbf{k}}}
	\]
	\[
	\times  \sum_{\mathbf{r}\mathbf{r}'}e^{i\mathbf{k}\cdot\left(\mathbf{r}-\mathbf{r}'\right)}\left(R^T_x(t)f^*_{\alpha x}(k_x)+R^T_y(t)f^*_{\alpha y}(k_y)\right)\hat{\sigma}_1\begin{pmatrix}
		0 & [D^{-1}]^A\\
		[D^{-1}]^R & [D^{-1}]^K
	\end{pmatrix}\hat{\sigma}_1\left(R_x(t')f_{\alpha x}(k_x)+R_y(t')f_{\alpha y}(k_y)\right)
	\]
	\begin{equation}\label{potentialaction}
		=\frac{n^2B^{'2}_0\hbar^2}{4mN}\int^{+\infty}_{-\infty}dt\sum_{\mathbf{k}}\sum_{\alpha\alpha'}\frac{k_{\alpha}k_{\alpha'}}{\omega_{\mathbf{k}}}\sum_{ij}R^q_i(t)\left(F_{ij}(\mathbf{k};\mathbf{r}\mathbf{r}')i\partial_t+G_{ij}(\mathbf{k};\mathbf{r}\mathbf{r}')\omega_{\mathbf{k}}+i0H_{ij}(\mathbf{k};\mathbf{r}\mathbf{r}')\right)R^{cl}_j(t)
	\end{equation}
	coming from $R_i\cdots R_j$ contributions. We have retained terms only up to linear order in $\mathbf{R}^q$, and here $F_{ij}(\mathbf{k};\mathbf{r}\mathbf{r}'), G_{ij}(\mathbf{k};\mathbf{r}\mathbf{r}'), H_{ij}(\mathbf{k};\mathbf{r}\mathbf{r}')$ are nonlocal (in $\mathbf{r},\mathbf{r}'$) and $\mathbf{k}$-dependent functions, the explicit expressions of which are
	\begin{equation}
		F_{ij}(\mathbf{k};\mathbf{r}\mathbf{r}')=\partial_iC_{\alpha'\mathbf{k}}(\mathbf{r})\partial_jC^*_{\alpha\mathbf{k}}(\mathbf{r})-\partial_iC^*_{\alpha'\mathbf{k}}(\mathbf{r})\partial_jC_{\alpha\mathbf{k}}(\mathbf{r}), G_{ij}(\mathbf{k};\mathbf{r}\mathbf{r}')=\partial_iC_{\alpha'\mathbf{k}}(\mathbf{r})\partial_jC^*_{\alpha\mathbf{k}}(\mathbf{r})+\partial_iC^*_{\alpha'\mathbf{k}}(\mathbf{r})\partial_jC_{\alpha\mathbf{k}}(\mathbf{r}), H_{ij}(\mathbf{k};\mathbf{r}\mathbf{r}')=-F_{ij}(\mathbf{k};\mathbf{r}\mathbf{r}')
	\end{equation}
	The next term is
	\[
	S_{kin}[\mathbf{R}]=  -\frac{\tau^2}{2}\frac{n^2B^{'2}_0\hbar^2}{2mN}\int dt\int^{+\infty}_{-\infty}dt'\int \sum_{\mathbf{k}}\sum_{\alpha\alpha'}\frac{k_{\alpha}k_{\alpha'}}{\omega_{\mathbf{k}}}
	\]
	\[
	\times  \sum_{\mathbf{r}\mathbf{r}'}e^{i\mathbf{k}\cdot\left(\mathbf{r}-\mathbf{r}'\right)}\left(\dot{R}^T_x(t)f^*_{\alpha x}(k_x)+\dot{R}^T_y(t)f^*_{\alpha y}(k_y)\right)\hat{\sigma}_1\begin{pmatrix}
		0 & [D^{-1}]^A\\
		[D^{-1}]^R & [D^{-1}]^K
	\end{pmatrix}\hat{\sigma}_1\left(\dot{R}_x(t')f_{\alpha x}(k_x)+\dot{R}_y(t')f_{\alpha y}(k_y)\right)
	\]
	\begin{equation}\label{kineticaction}
		=\frac{\tau^2n^2B^{'2}_0\hbar^2}{4mN}\int^{+\infty}_{-\infty}dt\sum_{\mathbf{k}}\sum_{\alpha\alpha'}\frac{k_{\alpha}k_{\alpha'}}{\omega_{\mathbf{k}}}\sum_{ij}R^q_i(t)\left(F_{ji}(\mathbf{k};\mathbf{r}\mathbf{r}')i\partial^3_t-G_{ij}(\mathbf{k};\mathbf{r}\mathbf{r}')\omega_{\mathbf{k}}\partial^2_t+i0H_{ji}(\mathbf{k};\mathbf{r}\mathbf{r}')\partial^2_t\right)R^{cl}_j(t)
	\end{equation}
	coming from $\dot{R}_i\cdots \dot{R}_j$ contributions,
	\[ S_{mixed}[\mathbf{R},\dot{\mathbf{R}}]=\frac{\tau}{2}\frac{n^2B^{'2}_0\hbar^2}{2mN}\int\int^{+\infty}_{-\infty}dtdt'\sum_{\mathbf{k}}\sum_{\alpha\alpha'}\frac{k_{\alpha}k_{\alpha'}}{\omega_{\mathbf{k}}}
	\]
	\[
	\times  \sum_{\mathbf{r}\mathbf{r}'}e^{i\mathbf{k}\cdot\left(\mathbf{r}-\mathbf{r}'\right)}\left(R^T_x(t)f^*_{\alpha x}(k_x)+R^T_y(t)f^*_{\alpha y}(k_y)\right)\hat{\sigma}_1\begin{pmatrix}
		0 & [D^{-1}]^A\\
		[D^{-1}]^R & [D^{-1}]^K
	\end{pmatrix}\hat{\sigma}_1\left(\dot{R}_x(t')f_{\alpha x}(k_x)+\dot{R}_y(t')f_{\alpha y}(k_y)\right)+\left(R_i\leftarrow\rightarrow \dot{R}_i\right)
	\]
	\begin{equation}\label{dissipativeaction}
		=0
	\end{equation}
	which vanishes within our linear in $\mathbf{R}^q$ approximation.

	Within this linear in $\mathbf{R}^q$ approximation, we can readily integrate this quantum fluctuation part of the Skyrmion collective coordinates and arrive at the equation of motion for the classical collective coordinates for the Skyrmion
	\begin{equation}
		\sum_{\mathbf{k}}\sum_{\alpha\alpha'}\frac{k_{\alpha}k_{\alpha'}}{\omega_{\mathbf{k}}}\sum_{ij}   F_{ji}(\mathbf{k};\mathbf{r}\mathbf{r}')i\tau^2\partial^3_tR^{cl}_j(t)-G_{ij}(\mathbf{k};\mathbf{r}\mathbf{r}')\tau^2\omega_{\mathbf{k}}\partial^2_tR^{cl}_j(t)+F_{ij}(\mathbf{k};\mathbf{r}\mathbf{r}')i\partial_tR^{cl}_j(t)+G_{ij}(\mathbf{k};\mathbf{r}\mathbf{r}')\omega_{\mathbf{k}}R^{cl}_j(t)=0
	\end{equation}
	which contains, in addition to the massive $\sim\partial^2_tR^{cl}_j(t)$ and conservative potential (confining force) term $\sim R^{cl}_j(t)$, also the higher order time derivative $\sim \partial^3_tR^{cl}_j(t)$ and dissipative friction $\sim\partial_tR^{cl}_j(t)$ terms. We can thus conclude that the magnetoelastic coupling also induces the mass term for the Skyrmion subject to a confining potential force, but at the expense of having dissipative term, which is not desired from the perspective of having non-dissipative dynamics.

	In our model, we have proposed a minimal coupling which is quadratic in the phonon displacement field. This is a simplified picture of the more accurate magnetoelastic coupling term quadratic in the phonon displacement field, which is allowed by the symmetry\cite{KaganovTsukerniksupp}
	\begin{equation}
		E_{me}\left[\mathbf{M},\mathbf{u}\right]=\sum_{\mathbf{r}}
		\left(L^{ik}_{lmrs}\frac{\partial \mathbf{M}(\mathbf{r})}{\partial r_{i}}\cdot\frac{\partial \mathbf{M}(\mathbf{r})}{\partial r_{k}}u_{lm}(\mathbf{r})u_{rs}(\mathbf{r})\right)
	\end{equation}
	where repeated indices are summed over, which is quadratic in the strain tensor (derivative of the $\mathbf{u}$) rather than the $\mathbf{u}$ itself. If compared to our minimal coupling, the effective coupling would be dependent on $\mathbf{k}$ instead of constant proportional to $\omega$. The resulting dynamics, which is non-dissipative and contains mass term in the equation of motion as predicted by our theory, will however remain to hold, but with much more complicated expression for the mass.
	
	To demonstrate this, we consider a simplest case where $L^{ik}_{lmrs}=L_0\delta^{ik}\delta_{lm}\delta_{rs}\delta_{lr}$, which gives
	\begin{equation}
		E_{me}\left[\mathbf{M},\mathbf{u}\right]=\sum_{\mathbf{r}}
		L_0\left(\frac{\partial \mathbf{M}(\mathbf{r})}{\partial r_{i}}\right)^2\left(\frac{\partial u_l(\mathbf{r})}{\partial r_l}\right)^2=-\frac{L_0}{N}\sum_{\mathbf{k},\mathbf{k}'}k_lk'_lu_l(\mathbf{k})u_l(\mathbf{k}')f_i(\mathbf{k},\mathbf{k}')
	\end{equation}
	where
	\begin{equation}
		f_i(\mathbf{k},\mathbf{k}')=\sum_{\mathbf{r}}\left(\frac{\partial \mathbf{M}(\mathbf{r})}{\partial r_{i}}\right)^2e^{-i\left(\mathbf{k}+\mathbf{k}'\right)\cdot\mathbf{r}}
	\end{equation}
	depending on the profile of the magnetization $\mathbf{M}(\mathbf{r})$, which is spatially nonuniform.

	In the presence of Skyrmion with collective coordinates $\mathbf{R}$, the coupling energy is found to be
	\begin{equation}
		E_{me}[\mathbf{R},\mathbf{u}]=\frac{L_0}{N}\sum_{\mathbf{k},\mathbf{k}'}k_lk'_lu_l(\mathbf{k})u_l(\mathbf{k}')\mathbf{R}\cdot\sum_{\mathbf{r}}\nabla\left[\left(\frac{\partial\mathbf{M}(\mathbf{r})}{\partial r_i}\right)^2e^{-i(\mathbf{k}+\mathbf{k}')\cdot\mathbf{r}}\right].
	\end{equation}
	We employ the transformation similar to Eq.(\ref{quantization}), 
	\begin{equation}
		u_l(\mathbf{k})=\sqrt{\frac{\hbar}{2m\omega_{\mathbf{k}}}}\left(b^{\dagger}_{l\mathbf{k}}+b_{l,-\mathbf{k}}\right)
	\end{equation}
	and retaining only the ``normal terms''(those of the form $b^{\dagger}b,bb^{\dagger}$), we arrive at
	\begin{equation}
		E^{normal}_{me}[\mathbf{R},b,b^{\dagger}]=\frac{\hbar L_0}{Nm\sqrt{\omega_{\mathbf{k}}\omega_{\mathbf{k}'}}}\mathbf{R}\cdot\sum_{\mathbf{k},\mathbf{k}'}b^{\dagger}_l(\mathbf{k})\mathbf{f}_{\mathbf{k},\mathbf{k}'}b_l(-\mathbf{k}')
	\end{equation}
	where the vector-valued kernel is given by
	\begin{equation}\label{vectorkernel}
		\mathbf{f}_{\mathbf{k},\mathbf{k}'}=k_lk'_l\sum_{\mathbf{r}}\nabla\left[\left(\frac{\partial\mathbf{M}(\mathbf{r})}{\partial r_i}\right)^2e^{-i(\mathbf{k}+\mathbf{k}')\cdot\mathbf{r}}\right],
	\end{equation}
	which can be shown to be a real-valued function for inversion-symmetric system; $\mathbf{f}^*_{\mathbf{k},\mathbf{k}'}=\mathbf{f}_{\mathbf{k},\mathbf{k}'}$. One can see that the magnetoelastic coupling energy between phonon and Skyrmion takes the form similar to our minimal coupling, except that now the coupling integral is much more complicated; it is no longer proportional to the total derivative of $b^{\dagger}b$, and so does not vanish upon integration over space. It is however clear that, following the same Keldysh Green's function approach detailed in the main text and the earlier sections of the supplementary materials, the magnetoelastic coupling quadratic in the phonon displacement field $\mathbf{u}$ will generate not only a mass term for the Skyrmion in its equation of motion, but also confining potential and dissipative friction terms.

	Detailed derivation is as follows. The Skyrmion-boson effective action in this case is given by
	\begin{equation}\label{SphiRsupp}
		S[\overline{\phi},\phi,\mathbf{R},\dot{\mathbf{R}}]=\int dt \sum_{\mathbf{k}\alpha}\overline{\phi}_{\alpha}(\mathbf{k},t)\hbar\left(i\partial_t-\omega_{\mathbf{k}}\right)\phi_{\alpha}(\mathbf{k},t)
		+\frac{1}{N}\int dt \sum_{\mathbf{k}\mathbf{k}'\alpha}\overline{\phi}_{\alpha}(\mathbf{k},t)\left(\frac{\hbar L_0}{m\sqrt{\omega_{\mathbf{k}}\omega_{\mathbf{k}'}}}\mathbf{R}\cdot\mathbf{f}_{\mathbf{k},\mathbf{k}'}-\hbar\omega_{\mathbf{k}}\tau\delta_{\mathbf{k},\mathbf{k}'}\dot{\mathbf{R}}\cdot i\nabla\right)\phi_{\alpha}(\mathbf{k}',t)
	\end{equation}  
	where, when compared to the corresponding action corresponding to our minimal coupling model; Eq.(13) in the main text, rather than a total derivative term, the first term in the parentheses of the second term in the above action involves a non-total derivative, proportional to $\mathbf{R}\cdot \mathbf{f}_{\mathbf{k},\mathbf{k}'}$. Looking at the form of the vector kernel $\mathbf{f}_{\mathbf{k},\mathbf{k}'}$ in Eq.(\ref{vectorkernel}), we can see that this kernel vanishes in the limit of small wave vector $k=|\mathbf{k}|,k'=|\mathbf{k}'|\rightarrow 0$ and or in the limit of large Skyrmion size, for which $\left(\partial\mathbf{M}/\partial r_i\right)^2$ is small, because the magnetization profile becomes very slowly varying with the spatial coordinate. The effective action of the coupled Skyrmion-phonon system then reduces to the effective action of our simplified model, with $\omega_{\mathbf{k}}=\omega$. Our simplified model that can describe the more accurate and realistic phonon system with the standard magnetoelastic coupling. Following the Keldysh approach detailed so far, integrating out the bosonic coherent states, one arrives at the effective action for the classical part of the Skyrmion collective coordinates
	\begin{equation}\label{effectiveactionMEcoupling}
		\delta S[\mathbf{R},\dot{\mathbf{R}}]=-\frac{1}{2}\int dt \int dt'\sum_{\mathbf{k}\mathbf{k}'}\hbar\omega _{\mathbf{k}}\left(\left(G^R[\tilde{G}^{-1}]^{q,cl}\right)^2+\left(G^A[\tilde{G}^{-1}]^{cl,q}\right)^2\right)
	\end{equation}
	where
	\begin{equation}
		[G^{-1}]^{A(R)}(t,t') = \hbar\left(i\partial_t-\omega_{\mathbf{k}}\mp i0\right)\delta(t-t')
	\end{equation}
	\begin{equation}\label{KeldyshGF}
		[G^{-1}]^K (t,t')=2i0 F(t,t')
	\end{equation}
	where $F(t,t')$ is a Hermitian matrix. On the other hand, the Green functions corresponding to bath-Skyrmion coupling action are
		\begin{equation}\label{pureclassicquantumparts}
			[\tilde{G}^{-1}]^{cl,cl}(t,t') = 0,  [\tilde{G}^{-1}]^{q,q}(t,t')= 2i 0 F(t,t') 
		\end{equation}
		\begin{equation}
			[\tilde{G}^{-1}]^{cl,q(q,cl)}(t,t') = \left(-2\frac{\hbar L_0}{m\sqrt{\omega_{\mathbf{k}}\omega_{\mathbf{k}'}}}\mathbf{R}_c\cdot\mathbf{f}_{\mathbf{k},\mathbf{k}'}+\tau\delta_{\mathbf{k},\mathbf{k}'}\dot{\mathbf{R}}_c\cdot\mathbf{k}\right)\delta(t-t')
		\end{equation}
	where we have retained only to zeroth order terms in $\mathbf{R}_q,\dot{\mathbf{R}}_q$. Substituting the Greens functions into Eq.(\ref{effectiveactionMEcoupling}) and using flatband dispersion $\omega_{\mathbf{k}}=\omega$, we obtain

	\begin{equation}
		\delta S=\int dt \sum_{\mathbf{k},\mathbf{k}'}\left[\frac{4\hbar^3L^2_0}{m^2\omega}\left(\mathbf{R}_c\cdot\mathbf{f}_{\mathbf{k}\mathbf{k}'}\right)^2-\frac{4\hbar^2L_0}{m}\left(\mathbf{R}\cdot\mathbf{f}_{\mathbf{k}\mathbf{k}'}\right)\tau\delta_{\mathbf{k}\mathbf{k}'}\dot{\mathbf{R}}_c\cdot\mathbf{k}+\hbar\omega\tau^2\delta_{\mathbf{k}\mathbf{k}'}\left(\dot{\mathbf{R}}_c\cdot\mathbf{k}\right)^2\right]
	\end{equation}
	which contains the confining potential energy term (the first term), dissipative friction term (second term), and the kinetic energy term (the third term). The last term is the only term that appears with our minimal coupling model, as presented in the main text. The magnetoelastic coupling thus produces more complicated effective action and equation of motion. But since $|\mathbf{f}_{\mathbf{k}\mathbf{k}'}|\rightarrow 0$ in the limit of large Skyrmion size, the result with magnetoelastic coupling reduce to that of our minimal coupling in this limit.

	\section{The Connection between Harmonic Oscillator Model and Phonon Model}
	
	Our main results as presented in the main text are obtained from a continuum model of harmonic oscillator lattice. This is a very simplified model of phonon system. In this simple model, the spectrum of the vibration is simply given by $\omega_{\mathbf{k}}=\omega$, where $\omega$ is the frequency of the harmonic oscillators. The spectrum is therefore dispersionless. This can be shown as follows. 
	
	In terms of ``normal coordinates'' defined in the discrete version $Q_{\mathbf{k}}=(1/\sqrt{N})\sum_{\mathbf{r}}\exp(i\mathbf{k}\cdot\mathbf{r})u(\mathbf{r})$ and their conjugate momenta $\Pi_{\mathbf{k}}=(1/\sqrt{N})\sum_{\mathbf{r}}\exp(i\mathbf{k}\cdot\mathbf{r})p(\mathbf{r})$ where $N$ is the number of lattice sites, the harmonic oscillator lattice Hamiltonian $H_B$ becomes
	\begin{equation}
		H_B=\frac{1}{2\rho}\sum_{\mathbf{k}}\Pi_{\mathbf{k}}\Pi_{-\mathbf{k}}+\frac{1}{2}\rho\sum_{\mathbf{k}}\omega^2_{\mathbf{k}}Q_{\mathbf{k}}Q_{-\mathbf{k}}
	\end{equation}
	where $\omega_{\mathbf{k}}=\omega$, where $\omega$ is the harmonic oscillator frequency, independent of the wave vector $\mathbf{k}$, thus mimicking the spectrum of an optical phonon mode, which is described in detail below.
	
	In standard model of lattice vibration \cite{BornHuangBookS}, a minimal model which hosts acoustic and optical phonon modes consists of a system of diatomic lattices; there are two atoms per unit cell, with masses $m$ and $m'$, separated by distance $s/2$ and so a distance $s$ between two consecutive atoms of the same type. With conservative potential $\phi(s/2)$ between the two atoms of the two types, solving the equations of motion gives two branches of spectrum with frequencies
	\begin{equation}
		\omega^2=\frac{\phi''\left(\frac{s}{2}\right)}{mm'}\left(m+m'-\sqrt{(m+m')^2-4mm'\sin^2\pi\eta}\right)
	\end{equation}
	for the acoustic branch and
	\begin{equation}
		\omega^2=\frac{\phi''\left(\frac{s}{2}\right)}{mm'}\left(m+m'+\sqrt{(m+m')^2-4mm'\sin^2\pi\eta}\right)
	\end{equation}
	for the optical branch, where $\eta$ is the wave vector with Brillouin zone $-1/2\leq \eta\leq 1/2$. The two branches of phonon dispersion are illustrated in Fig.~\ref{fig:PhononDispersion}. We have taken a harmonic potential $\phi(x)=Kx^2/2$ for simplicity. One can see that the optical branch is very flat, the flatness of which depends on the mass ratio of the diatom; the more different the masses of the two atoms per unit cell are, the flatter the optical phonon spectrum will be. Our simple model with independent harmonic oscillators thus mimic the optical phonon mode of real phonon system. Therefore, from this point onward, we will focus on the optical phonon mode.
	
	\begin{figure}
		\includegraphics[angle=0,origin=c, scale=1.0]{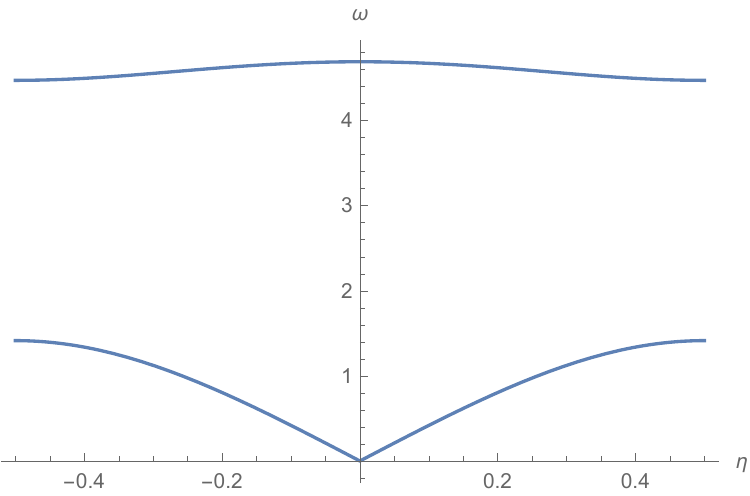}
		\caption{
			Typical profile of optical phonon spectrum (upper) and acoustic phonon spectrum (lower).}
		\label{fig:PhononDispersion}
	\end{figure}
	
	To analyze further this connection, we recast the phonon spectrum in terms of effective mass of the diatomic system,
	\begin{equation}
		\mu = \frac{mm'}{m+m'}
	\end{equation}
	which gives
	\begin{equation}
		\omega^2=K\left(\frac{1}{\mu}+\sqrt{\frac{1}{\mu^2}-4\frac{\mu}{M}\sin^2\pi\eta}\right)
	\end{equation}
	where $M=m+m'$ is the total mass of the diatomic system. Now, consider the limit of very different masses; $m\gg m'$, we have $\mu\simeq m'\ll M$. In this limit, we have
	\begin{equation}
		\omega^2\simeq \frac{2K}{m'}
	\end{equation}
	independent of the wave vector $\eta$. Our simple model Hamiltonian of harmonic oscillators with 2D mass density $\rho$ thus corresponds to phonon system with diatomic mass $\mu\simeq m'$ per unit cell, of area $A=a^2$. To conclude, independent harmonic oscillator system as used in our model, with mass $m_0=\rho a^2$ and frequency $\omega$, corresponds to the optical mode of phonon system with frequency $\omega^2=2K/m'$ where $K$ is the effective spring constant and $m'$ the lighter mass of the two atoms per unit cell.
	
	\textcolor{black}{Now, we give a schematic outline on how to extend our theory to full phonon model which gives proper dispersion relation covering both optical and acoustic phonons. A minimal model in the real space lattice would have to be a two-atom per unit cell lattice model, which would translate to two-band model in reciprocals space. The starting real space lattice model can be rather complicated but the resulting reciprocal space Hamiltonian can be written as
		\begin{equation}
			H_B=\sum_{\mathbf{k}}\sum_{a=\mathrm{ac},\mathrm{op},\alpha=x,y} \hbar\omega_a(\mathbf{k})\left(b^{\dag}_{a,\alpha}(\mathbf{k})b_{a,\alpha}(\mathbf{k})+\frac{1}{2}\right)
		\end{equation}
		where $a=\mathrm{ac},\mathrm{op}$ refers to the acoustic, optical phonon respectively and $\omega_{a}(\mathbf{k})$ represents the corresponding dispersion relation of the respective phonon $a=\mathrm{ac},\mathrm{op}$ and we have implicitly assumed isotropic case for which the dispersion is the same along $x$ and $y$ directions. The Skyrmion couples equally to the two types of phonon (optic and acoustic), as does the standard magnetoelastic coupling. The derivation proceeds following the line detailed in the previous section, where the effective action would be written in the reciprocal space, thus accommodating the $\mathbf{k}$-dependence of the acoustic phonon dispersion relation. Integrating out the phonon degree of freedom would give an effective action for the Skyrmion collective coordinates, which now involve $\mathbf{k}$-dependent effective coupling in terms of the acoustic phonon dispersion relation $\omega(\mathbf{k})$. The resulting Skyrmion mass and cyclotronic frequency would receive contributions from all values of wave vector $\mathbf{k}$. However, in practice, acoustic phonon of a given wave vector $\mathbf{k}_0$ and corresponding frequency $\omega(\mathbf{k}_0)$ can be excited at resonance by some driving field. Varying the resonance frequency can thus vary the Skyrmion effective mass and corresponding cyclotronic frequency, thus realizing the coveted controllability of the dynamics envisioned in this work. It is to be noted that in this line of derivation, optical phonon always comes along and accompanies the acoustic phonon. The total effect of the phonons on the Skyrmion dynamics would sum up the contributions of these two types of phonons. The relative strength of the two contributions would after all depend on the operating frequency that can be controlled by the resonance condition; the resonance frequency and the type of exciting field; e.g. acoustic driving field for acoustic phonon and optical driving field for optical phonon. In the proposed candidate material NiGa$_2$S$_4$, the phonon modes relevant to the above description are the in-plane $E_g$ modes, consisting of three modes, one of them is very weak. The other two are Raman active and can be approximately described by the two-band phonon model as sketched above.}

\end{widetext}


\begin{thebibliography}{1}
\bibitem{Note} One can also consider a direct coupling of the form $\lambda (\mathbf{u}\cdot\mathbf{R})^2$, at the cost of introducing a rather arbitrary coupling parameter $\lambda$. This will eventually give rise to a coupling quadratic in bosonic operator, but also quadratic order in the Skyrmion displacement vector $\mathbf{R}$. A linear coupling of the form $\lambda \mathbf{u}\cdot\mathbf{R}$ is not allowed as it breaks the reflection symmetry $\mathbf{u}\rightarrow -\mathbf{u}$, while it is clear the $\pm \mathbf{u}$ displacements for the harmonic oscillator must have the same energy. Adding this direct linear coupling to the harmonic oscillator Hamiltonian Eq.(\ref{BathHamiltonian}), we can rewrite it as $\rho\dot{\mathbf{u}}^2/2+\rho\omega^2\mathbf{u}^2/2+\lambda \mathbf{u}\cdot\mathbf{R}=\rho\dot{\mathbf{u}}^2/2+\rho\omega^2(\mathbf{u}-\mathbf{u}_0)^2/2-(\lambda^2/(2\rho\omega^2))\mathbf{R}^2$ where $\mathbf{u}_0=-(\lambda/(\rho\omega^2))\mathbf{R}$, showing that this coupling makes the system unstable because its energy becomes unbounded from below; the displacement vectors $\mathbf{u}$ and $\mathbf{R}$ will both tend to grow infinitely large in magnitudes in order to minimize the energy to arbitrarily large negative value. On the other hand, the quadratic coupling of the form $\lambda\left(\mathbf{u}\cdot\mathbf{R}\right)^2$ simply amounts to anisotropic renormalization of the oscillator frequency $\omega\rightarrow\omega_{\alpha}$ where $\omega^2_{\alpha}=\omega^2+\left(2\lambda/\rho\right)R^2_{\alpha}$ where $\alpha=x,y$. Within perturbative theory applicable at high frequency $\lambda R^2_{\alpha}/(\rho\omega^2)\ll 1$, this coupling produces a potential energy term quartic in $R$ (that is, $\sim R^4$) rather than a kinetic energy term as in Eq.(\ref{SkyrmionActionCorrection}).
\bibitem{NoteNotation}In the rest of this article, the length of the spin $|\mathbf{S}|$ is normalized to unity. The spin vector, whenever appears, is denoted by $\mathbf{S}$ while the action is denoted by $S$ with appropriate subscript or superscript that specifies what action it represents.
\end{thebibliography}
\end{document}